\documentclass[useAMS,usenatbib]{mn2e}
\usepackage{epsfig}
\usepackage{graphicx}
\usepackage{epstopdf}
\DeclareGraphicsExtensions{.ps}
\usepackage{times}
\usepackage{natbib}
\usepackage{amssymb}
\usepackage{amsmath}
\usepackage{tabularx}
\usepackage{amsfonts,amsmath,amssymb}
\usepackage{txfonts}
\usepackage{graphicx}
\usepackage{color}
\usepackage{flushend}
\usepackage{enumitem}
\usepackage{tabularx}
\usepackage{color}
\newif\ifAMStwofonts
\AMStwofontstrue
\newcommand {\aplt} {\ {\raise-.5ex\hbox{$\buildrel<\over\sim$}}\ } 
\newcommand{\redmapper}{RM}
\newcommand{\be}{\begin{equation}}
\newcommand{\ee}{\end{equation}}
\newcommand{\bes}{\begin{equation*}}
\newcommand{\ees}{\end{equation*}}
\newcommand{\ba}{\begin{eqnarray}}
\newcommand{\ea}{\end{eqnarray}}
\newcommand{\bas}{\begin{eqnarray*}}
\newcommand{\eas}{\end{eqnarray*}}
\newcommand{\brr}{\begin{array}}
\newcommand{\err}{\end{array}}
\newcommand{\bc}{\begin{center}}
\newcommand{\ec}{\end{center}}
\newcommand{\bit}{\begin{itemize}}
\newcommand{\eit}{\end{itemize}}

\newcommand{\msun}{\,{\rm M}_\odot}
\newcommand{\msunh}{\,h_{70}^{-1}{\rm M}_\odot}

\newcommand{\degs}{\mbox{$\textrm{deg}^2$ }}

\newcommand{\lam}{\mbox{$\lambda$}}

\newcommand{\nvt}{\mbox{$\textrm{N}_\textrm{vt}$}}

\newcommand{\vel}{\,{\rm km\,s^{-1}}}

\newcommand{\mincir}{\raise
  -2.truept\hbox{\rlap{\hbox{$\sim$}}\raise5.truept \hbox{$<$}\ }}
\newcommand{\magcir}{\raise
  -2.truept\hbox{\rlap{\hbox{$\sim$}}\raise5.truept \hbox{$>$}\ }}
\newcommand{\siml}{\raise
  -2.truept\hbox{\rlap{\hbox{$\sim$}}\raise5.truept \hbox{$<$}\ }}
\newcommand{\simg}{\raise
  -2.truept\hbox{\rlap{\hbox{$\sim$}}\raise5.truept \hbox{$>$}\ }}

\newcommand{\rvir}{$R_{500}$}
\newcommand{\mvir}{$M_{500}$}

\title[]{Constraints on the Richness-Mass Relation and the Optical-SZE
  Positional Offset Distribution for SZE-Selected Clusters}
\author[A. Saro, et al.]{A.~Saro$^{1,2}$, S.~Bocquet$^{1,2}$, E.~Rozo$^{3}$, B.~A.~Benson$^{4,5,6}$, J.~Mohr$^{1,2,7}$, E.~S.~Rykoff$^{8,9}$, 
\newauthor M.~Soares-Santos$^{4}$, L.~Bleem$^{5,10}$, S.~Dodelson$^{4,5,6}$, P.~Melchior$^{11,12}$
% ...
\\ \smallskip \textit{Affiliations are listed at the end of the paper}}

\begin{document}
\pdfpageheight 11.7in
\pdfpagewidth 8.3in
\date{Accepted ???. Received ???; in original form ???}   
\maketitle                                                 
\begin{abstract} 
We cross-match galaxy cluster candidates selected via their Sunyaev-Zel'dovich effect (SZE) signatures in 129.1~\degs of the South Pole Telescope 2500d SPT-SZ survey with optically identified clusters selected from the Dark Energy Survey (DES) science verification data.  We identify 25 clusters between $0.1 \lesssim z \lesssim 0.8$ in the union of the SPT-SZ and redMaPPer (RM) samples.  RM is an optical cluster finding algorithm that also returns a richness estimate for each cluster. We model the richness $\lambda$-mass relation with the following function  $\langle \ln \lambda|M_{500} \rangle \propto B_\lambda\ln M_{500} + C_\lambda\ln E(z)$ and use SPT-SZ cluster masses and RM richnesses $\lambda$ to constrain the parameters. We find $B_\lambda= 1.14^{+0.21}_{-0.18}$ and $C_\lambda=0.73^{+0.77}_{-0.75}$.  The associated scatter in mass at fixed richness is $\sigma_{\ln M|\lambda} = 0.18^{+0.08}_{-0.05}$ at a characteristic richness $\lambda=70$.  We demonstrate that our model provides an adequate description of the matched sample, showing that the fraction of SPT-SZ selected clusters with RM counterparts is consistent with expectations and that the fraction of RM selected clusters with SPT-SZ counterparts is in mild tension with expectation.  We model the optical-SZE cluster positional offset distribution with the sum of two Gaussians, showing that it is consistent with a dominant, centrally peaked population and a sub-dominant population characterized by larger offsets. We also cross-match the RM catalog with SPT-SZ candidates below the official catalog threshold significance $\xi=4.5$, using the RM catalog to provide optical confirmation and redshifts for additional low-$\xi$ SPT-SZ candidates. In this way, we identify 15 additional clusters with $\xi\in [4,4.5]$ over the redshift regime explored by RM in the overlapping region between DES science verification data and the SPT-SZ survey.
\end{abstract}

\section {Introduction}
\label{sec:introduction}
Clusters of galaxies were first identified as over-dense regions in
the projected number counts of galaxies
\citep[e.g.,][]{Abell58,ZW68.1}.  
Nowadays, clusters are also regularly identified through their X-ray emission
\citep[e.g.,][]{gioia90,vikhlinin98,boehringer00,pacaud07,suhada12}
and at millimeter wavelengths through their Sunyaev-Zel'dovich
effect (SZE) signatures \citep{Sunyaev72}.  Large, homogeneously selected samples of clusters are useful for both cosmological and astrophysical studies, and such samples have recently begun to be produced using SZE selection 
\citep[][]{staniszewski09,hasselfield13,planck15clusters,bleem15}.
There is a longer history of large cluster samples selected from optical and near
infrared photometric surveys \citep[e.g.,][and references therein]{gladders00,koester07a,eisenhardt08,menanteau10,hao10,wen12,rykoff14,bleem14b,ascaso14}, and even larger samples will soon be available from ongoing and future surveys like the Dark Energy Survey \citep[DES,][]{des}\footnote{http://www.darkenergysurvey.org}, KiDS \citep{kids}, Euclid \citep{euclid} and LSST \citep{lsst}.

Reliable estimates of galaxy cluster masses play a key role in both cosmological and astrophysical cluster studies. First, the abundance of galaxy clusters as a function of mass is a well-known cosmological probe
\citep[][and many others]{white93b,bartlett94,eke98,viana99,borgani01, vikhlinin09, rozo10, mantz10,allen11, benson11,bocquet15}. Second, accurate estimates of cluster masses are crucial in disentangling environmental effects from the secular evolution processes shaping galaxy formation
\citep{mei09,zenteno10,muzzin12}.

In this paper, we calibrate the richness-mass relation for SZE-selected galaxy clusters detected in the DES science verification data (SVA1) using the redMaPPer (Rykoff et al. 2014) cluster-finding algorithm.  Specifically, we study the clusters detected via their SZE signatures in the South Pole Telescope SPT-SZ
cluster survey \citep[][hereafter B15]{bleem15} that are also present in the redMaPPer catalog. We also study the distribution of offsets between the SZE derived centres and the associated optical centres, properly including the SZE positional uncertainties. Finally, we demonstrate our ability to push to even lower candidate significance within the SPT-SZ candidate catalog by taking advantage of the contiguous, deep, multiband imaging available through DES.  In this respect, our study points towards the combined use of DES and SPT datasets to provide highly reliable extended SZE-selected cluster samples.  We note that historically the optical follow-up of SPT selected clusters was the original motivation for proposing DES.

The plan of the paper is as follows. In Section~\ref{sec:data} we describe the galaxy cluster catalogs and the matching metric we use in this work. Section~\ref{sec:method} describes the method we adopt to
calibrate the SZE-mass and richness-mass relations. Our results are presented in Section \ref{sec:results}. Section~\ref{sec:conclusions} contains a discussion of our findings and our conclusions.  In the Appendix, we provide a preliminary analysis of a cluster sample created using an independent cluster finding algorithm --- the Voronoi Tessellation (VT) cluster finder --- which helps to highlight areas where the VT algorithm can be
improved.  Throughout this work, we adopt $\Omega_M= 0.3$, $\Omega_\Lambda =0.7$, $H_0 = 70 $ $\vel$ Mpc$^{-1}$, and $\sigma_8 =0.8 $. Cluster masses are defined within \rvir, the radius within which the density is 500 times the critical density of the Universe. Future analyses will include the dependence of the derived scaling relation parameters on the adopted cosmology by simultaneously fitting for cosmological and scaling-relation parameters \citep[e.g.][]{mantz10,rozo10,bocquet15}.

\section{Cluster Sample}
\label{sec:data}

\subsection{SPT-SZ Cluster Catalog}
\label{sec:spt}

The SPT-SZ galaxy cluster sample used in this analysis has been selected via the cluster thermal SZE signatures in a point-source masked-region of 2365~\degs of the 2540~\degs (2500d) SPT-SZ survey using 95~GHz and 150~GHz data. Typical instrumental noise is approximately 40~(18) $\mu {\rm K_{CMB}}$-arcmin and the beam FWHM is 1.6~(1.2) arcmin for the 95~(150)~GHz maps. A multi-frequency matched filter is used to extract the cluster SZE signal in a manner designed to optimally measure the cluster signal given knowledge of the cluster profile, the SZE spectrum and the noise in the maps \citep{haehnelt96,melin06}. The cluster gas profiles are assumed to be described by a
projected isothermal $\beta$ model \citep[]{cavaliere76} with $\beta=1$.  Note that, as discussed in \citet{vanderlinde10}, the resulting SPT-SZ candidate catalogs are not sensitive to this assumption.  The adopted model provides a SZE temperature decrement that is maximum at the cluster centre and weakens with separation $\theta$ from the cluster centre as: 
\be \Delta T(\theta) = \Delta T_0 [1 +
  (\theta/\theta_c)^2]^{-1}, 
\ee 
where $\Delta T_0$ is the central value and $\theta_c$ is the core radius. We adopt 12 different cluster profiles linearly spaced from $\theta_c= 0.25$ to 3 arcmin \citep[][B15]{vanderlinde10,reichardt13}.  For each cluster, the maximum signal-to-noise across the 12 filtered maps is denoted as $\xi$.  The SPT-SZ cluster candidates with $\xi>4.5$ have been previously published in B15.

\subsection{DES Optical Cluster Catalogs}
\label{sec:des}

The DES Science Verification Data (DES-SVA1) that overlap SPT have been used to produce optically selected catalogs of clusters.  In Section~\ref{subsec:des} we describe the acquisition and preparation of the DES-SVA1 data, and in Section~\ref{subsec:redmapper_clusters} we describe the production of the redMaPPer cluster catalog used in the primary analysis.  We remind the reader that in Appendix~\ref{sec:vt_analysis} we present results of a preliminary analysis of the VT cluster catalog.

\subsubsection{DES-SVA1 Data}
\label{subsec:des}
The DES-SVA1 data include imaging of $\sim300$~\degs over multiple disconnected fields \citep[]{melchior15, sanchez14,banerji14}, most of which overlap with the SPT-SZ survey. The DES-SVA1 data were acquired with the Dark Energy Camera \citep{diehl12,flaugher12,flaugher15} over 78 nights, starting in Fall 2012 and ending early in 2013 with depth comparable to the nominal depth of the full DES survey (Rykoff et al., in preparation).

Data have been processed through the DES Data Management (DESDM, \citealt{desai12}) pipeline that is an advanced version of development versions described in several publications \citep{ngeow06,mohr08,mohr12}.  The data were calibrated in several stages leading to a \textit{Gold} catalog of DES-SVA1 galaxies (Rykoff et al., in preparation). The \textit{Gold} catalog covers $\sim250$~\degs and is optimized for extragalactic science. In particular it masks regions south of declination $\delta=-61^\circ$, avoiding the Large Magellanic Cloud and its high stellar densities. Furthermore, the footprint is restricted to the regions where we have coverage in all four bands.

\subsubsection{redMaPPer Cluster Catalog}
\label{subsec:redmapper_clusters}

The red-sequence Matched-Filter Probabilistic Percolation (redMaPPer, hereafter RM) algorithm is a cluster-finding algorithm based on the richness estimator of \citet{rykoff12}.  RM has been applied to photometric data from the Eighth Data Release (DR8) of the Sloan Digital Sky Survey \citep[SDSS,]{aihara11} and to the SDSS Stripe 82 coadd data \citep{annis14}, and has been shown to provide excellent photometric redshifts, richness estimates that tightly correlate with external mass proxies, and very good completeness and purity \citep{rozo14b,rozo14d, rozo14a}. We refer the reader to the paper by \cite{rykoff14} for a detailed description of the algorithm. Here, we briefly summarize the most salient features. 

We employ an updated version of the algorithm (v6.3.3), with improvements summarized in \citet{rozo14b}, \citet[in preparation]{rozo15}, and \citet[in preparation]{rykoff15}. RM calibrates the colour of red-sequence
galaxies using galaxy clusters with spectroscopic redshifts. RM uses this information to estimate the membership probability of every galaxy in the vicinity of a galaxy cluster. The richness $\lambda$ is
thus defined as the sum of the membership probabilities ($p_{\mathrm{RM}}$) over all galaxies:
\begin{equation}
\lambda = \sum p_\textrm{RM}. 
\end{equation}
In addition to the estimate of membership probabilities, the RM centering algorithm is also probabilistic. The centering probability $P_\textrm{cen}$ is a likelihood-based estimate of the probability that the galaxy under consideration is a central galaxy.  The centering likelihood includes the fact that the photometric redshift of the central galaxy must be consistent with the cluster redshift, that the central galaxy luminosity must be consistent with the expected luminosity of the central galaxy of a cluster of the observed richness, and that that the galaxy density on a 300 kpc scale consistent with the galaxy density of central galaxies.  The centering probability further accounts for the fact that every cluster has one and only one central galaxy, properly accounting for the relevant combinatoric factors. 
These probabilities have been tested on SDSS DR8 data using X-ray selected galaxy clusters, and have been shown to produce cluster centres that are consistent with the X-ray centres \citep{rozo14b}. 

The DES-SVA1 RM catalog was produced by running on a smaller footprint than that for the full SVA1 \textit{Gold} sample. In particular, we restrict the catalog to the regions where the $z$-band $10\sigma$ galaxy limiting magnitude is $z>22$. In total, we use $148\,\mathrm{deg}^2$ of DES-SVA1 imaging, with $129.1\,\mathrm{deg}^2$ overlapping the SPT-SZ footprint.  In this area, the largest fraction ($124.6\,\mathrm{deg}^2$) is included in the so called DES-SVA1 SPT-E field. The final catalog used in this work consists of 9281 clusters with $\lambda > 5$ and redshifts in the range $0.1<z<0.9$. Due to the varying depth of the DES-SVA1 catalog, RM produces a mask that determines the maximum redshift of the cluster search at any given location in the survey.  As an example, the effective area in the SPT-E region at the highest redshift ($z>0.85$) is only $\sim30\,\mathrm{deg}^2$.  In addition to the cluster catalog, the RM algorithm also uses the survey mask to produce a set of random points with the same richness and redshift distribution as the clusters in the catalog.  The random points take into account the survey geometry and the physical extent of the clusters, and as with the clusters, only includes points that have $<20\%$ of the local region masked \citep[see][]{rykoff14}.  

\subsection{Catalog Matching}
\label{sec:catalog_match}
We cross-match the SPT-SZ catalog with the RM optical cluster catalog following the method of \citet{rozo14a}.  First, we sort the SPT-SZ clusters to produce a list with decreasing SZE observable $\xi$, and we sort the RM catalog to produce a list with decreasing richness.  Second, we go down the SPT-SZ sorted list, associating each SPT cluster candidate with the richest RM cluster candidate whose centre lies within 1.5 \rvir \, of the SZE centre.  Third, we remove the associated RM cluster from the list of possible counterparts when matching the remaining SPT-selected clusters.  

\rvir \, is first computed assuming the redshift of the optical counterpart and using the SZE-mass scaling relation parameters adopted in B15.  We subsequently check that our sample does not change when adopting our best fitting scaling relation parameters (see Section \ref{sec:sze_mass}).

\begin{figure}
\hbox to \hsize{
    \includegraphics[width=85mm]{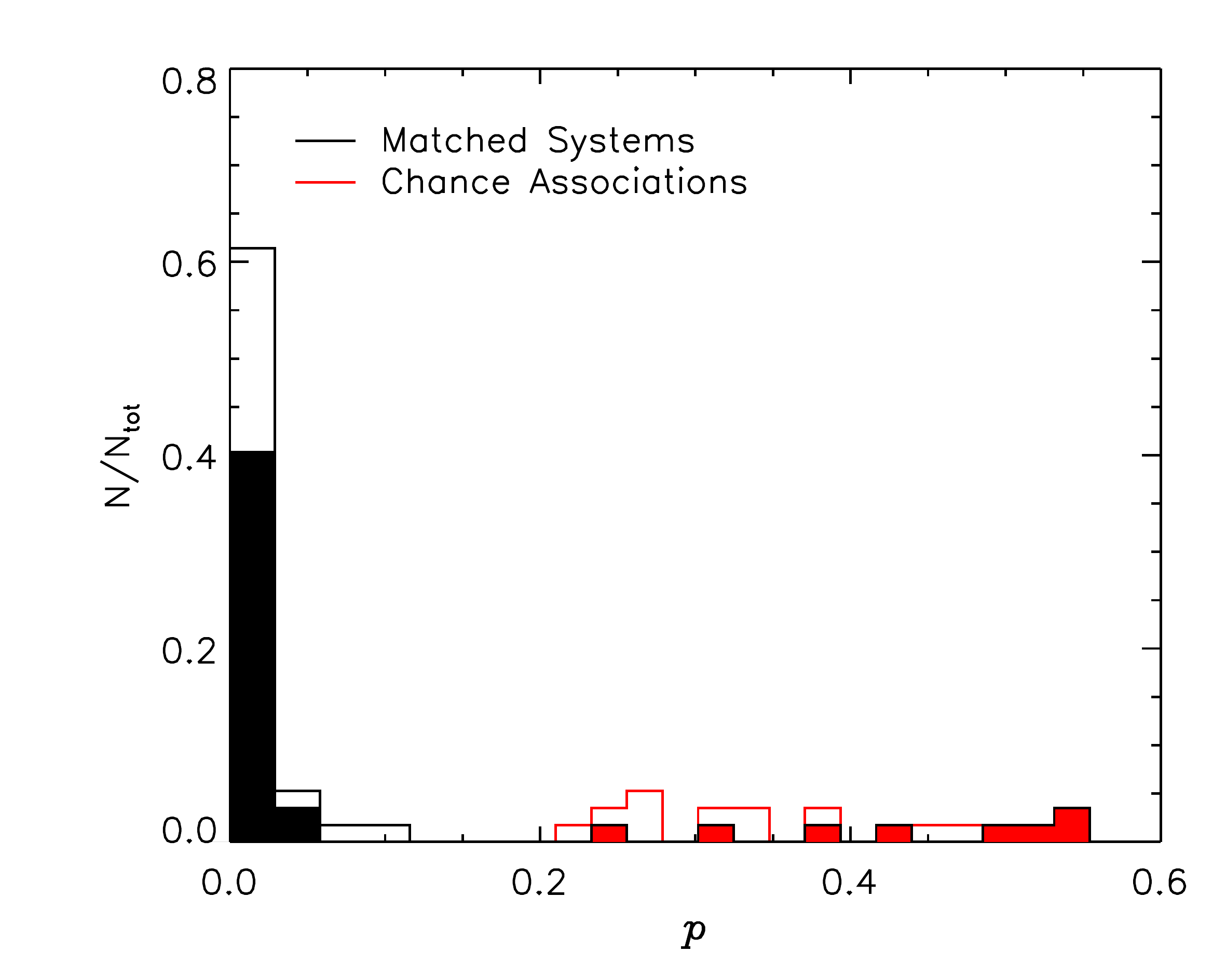}\hfil
 }
\caption{Distribution of probability $p$ of chance associations for the $\xi>4.5$ (filled) and $\xi>4$ (empty) samples. Black lines show the sample used in this work, red lines show the sample rejected (Section \ref{sec:catalog_match}).
\label{fig:probability}}
\end{figure}

To test the robustness of our matching algorithm against chance associations, we first perform the above described procedure on a sample of randomly generated RM clusters as described in the previous Section. Positions of clusters in this randomly generated sample do not correlate with the positions of the SPT-SZ clusters. Using an ensemble of $10^4$ random catalogs we measure the distribution of richness in chance associations for each SPT-SZ candidate. We then apply the algorithm to match the real RM cluster catalog with the SPT-SZ candidate list where $\xi>4.5$. The distribution of probabilities $p$ of chance associations estimated for each cluster candidate using the randomly generated samples is shown in Figure~\ref{fig:probability} (filled histogram).

\begin{figure*}
\hbox to \hsize{
    \includegraphics[width=85mm]{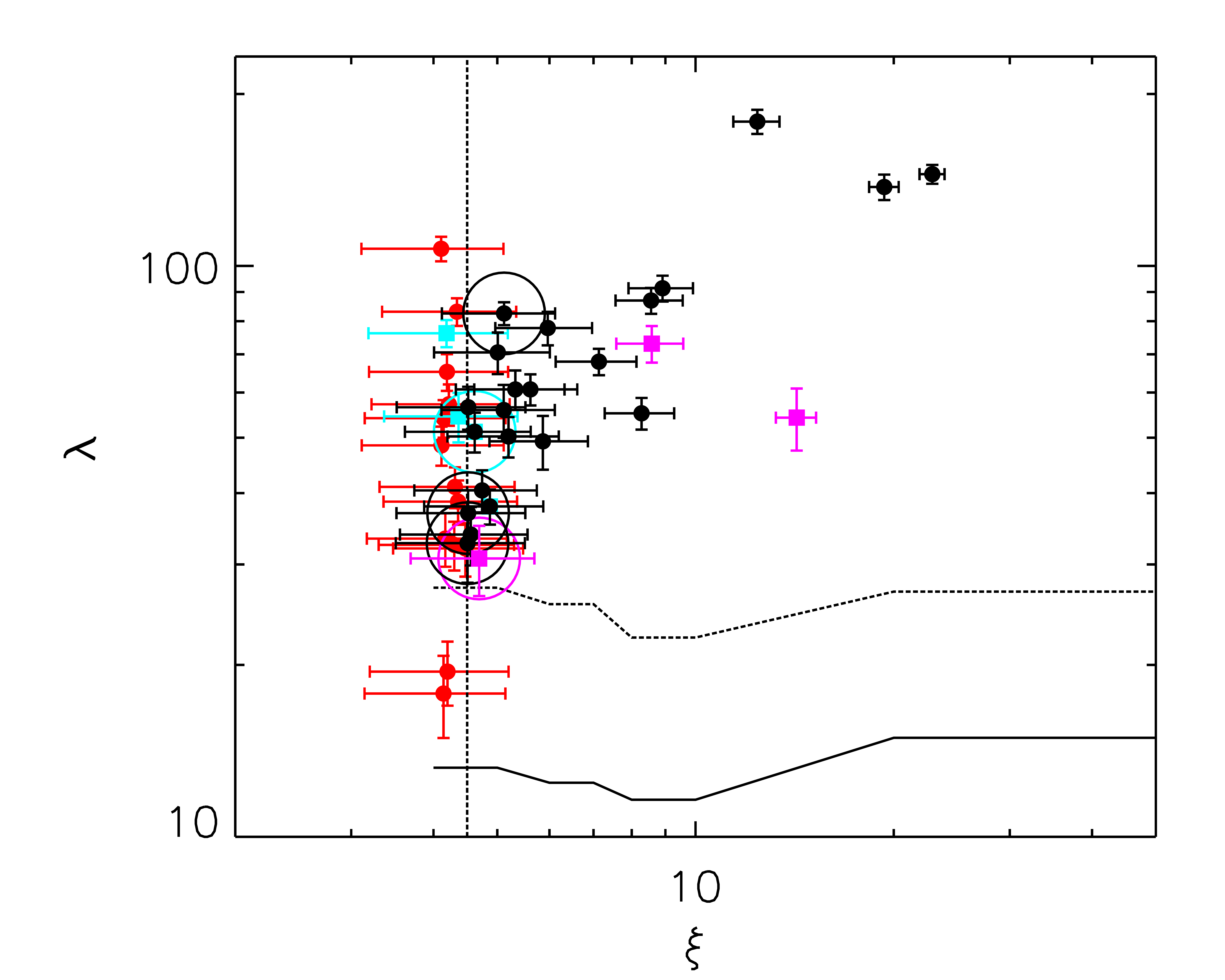} 
    \includegraphics[width=85mm]{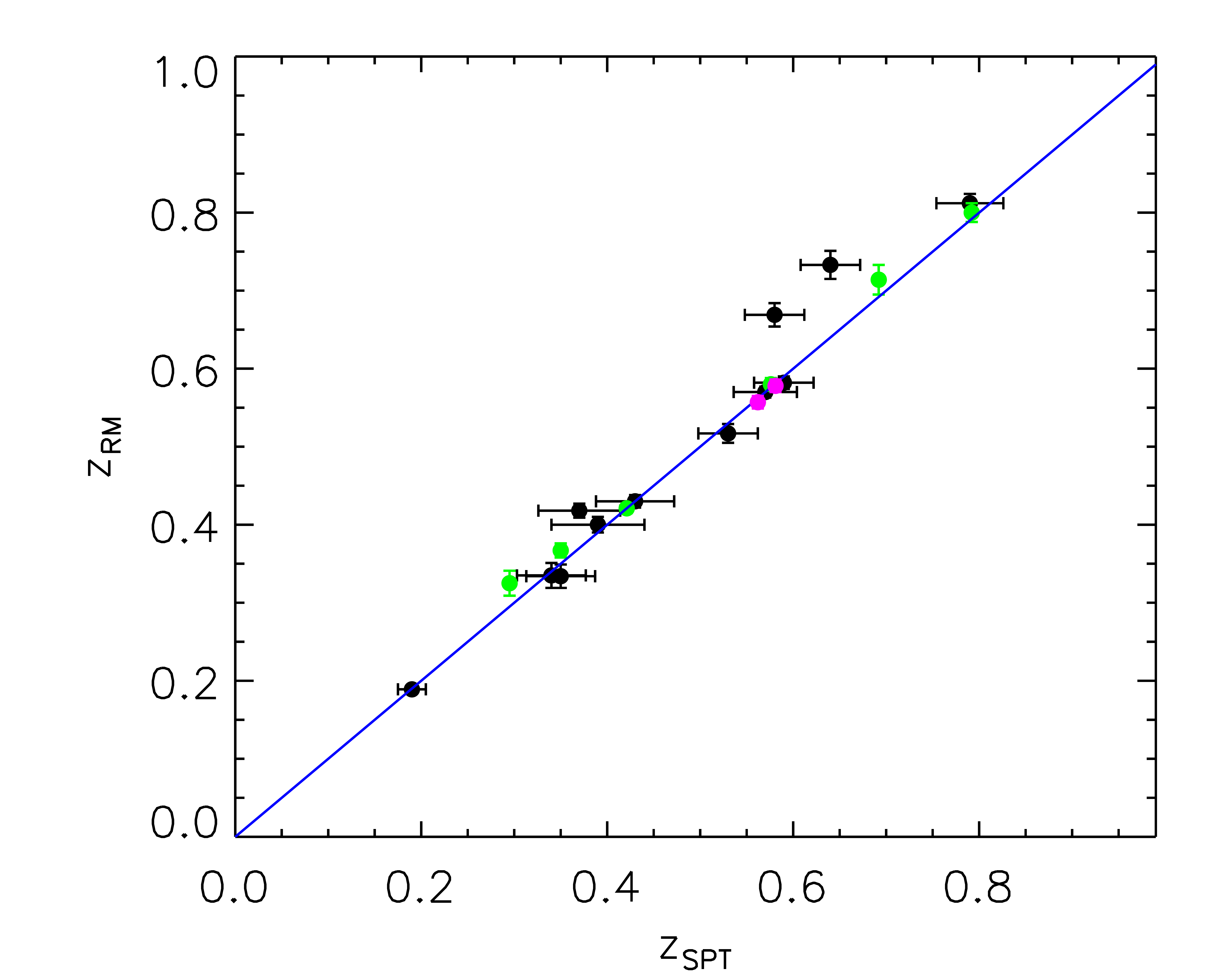} 
\hfil} 
    \vskip-0.1cm
\caption{\textit{Left panel:} Richness as a function of the SZE
  significance $\xi$ for the matched cluster sample. SPT-SZ candidates
  with $\xi<4.5$ (vertical line) are shown in red.  Clusters at
  $z<0.25$ (cyan) and clusters with miscentering due to a high masked fraction (magenta) are not used in the richness analysis.  Large circles indicate the newly
  confirmed SPT-SZ candidates with $\xi>4.5$.  Solid and dashed lines
  represent the upper 84 and 95 percentiles in richness of chance associations of SPT-SZ candidates and clusters from the randomly generated RM catalog.
  \textit{Right panel:} The estimated redshift for the RM sample as a
  function of SPT redshifts as presented in B15 from independent
  optical follow-up data. SPT-SZ candidates with spectroscopic redshift
  are shown in green. Magenta symbols are the same as in the left panel.  }
\label{fig:match}
\end{figure*}

In Figure~\ref{fig:match} (left panel) we show the resulting 84\% and 95\% confidence limits (solid and dotted lines, respectively) in the richness distribution of the chance associations as a function of the SPT-SZ observable $\xi$. This test allows us to estimate the probability of chance superposition for each SPT-SZ cluster candidate. As detailed below, we use this information to determine whether or not to include particular matches for further analysis.  As Figure~\ref{fig:match} shows, this filtering of the matched sample then ensures that chance superpositions are playing no more than a minor role even at $4<\xi<4.5$.

Within the DES-SVA1 region explored by RM there are 36 such SPT-SZ cluster candidates.  Using information on the contamination fraction at $\xi>4.5$ of the SPT-SZ candidate list and on the redshift distribution of the confirmed cluster candidates \citep[][B15]{song12}, we expect $\sim 9$ of these candidates to be noise fluctuations and $\sim 80\%$ of the real clusters to lie at $z<0.8$. Therefore, assuming the optical catalog is complete, the expected number of real cluster matches is $\sim22$. Similarly, one can estimate an expected number of real cluster matches (23.6) by scaling the total number of confirmed clusters in B15 below $z=0.8$ (433) in the 2365~\degs of the SPT-SZ survey by the DES-SVA1 area overlapping SPT-SZ that has been processed with the RM cluster finder (129.1~\degs).

The actual number of matches to the SPT-SZ candidates is 33. Eight systems are foreground, low-richness RM clusters that have been erroneously associated with SPT-SZ candidates with either previously measured redshifts (four systems) or lower limits estimated in B15 (the remaining four candidates) that are at $z \gtrsim 0.8$; these systems are either noise fluctuations or real clusters that are at redshifts too high for them to be detected by \redmapper. In fact, all of these systems have a probability $p$ of chance associations estimated from the randomly generated sample that is $p>16\%$. Therefore, we remove these matches from the sample. This leaves 25 SPT-SZ candidates at $\xi > 4.5$ that have RM counterparts. We expect less than one false associations within this sample of 25 candidates. The associated optical richness
as a function of the SPT-SZ significance is shown in the left panel of Figure~\ref{fig:match}. This number is somewhat larger but statistically consistent with the expected number of matches presented above. All 22 of the SPT-SZ confirmed clusters presented in B15 that lie at redshifts where they could be detected by RM are in this matched sample. According to our matching metric (which differs from the approach in B15), there are also three unconfirmed SPT-SZ candidates (i.e., candidates without identified optical counterparts in the B15 analysis) that have RM counterparts: SPT-CL J0502-6048, SPT-CL J0437-5307 and SPT-CL J0500-4551.  The newly confirmed clusters are highlighted with large circles in Figure \ref{fig:match}. 

Of the 25 SPT-SZ candidates with robust RM counterparts, we use 19 of them to calibrate the RM richness-mass relation. Six clusters are excluded from the analysis for the following reasons.  Clusters with estimated redshift $z<0.25$ in the SPT-SZ catalog from B15 are highlighted in cyan in the left panel of Figure~\ref{fig:match}. Because the $\xi$-mass relation is robust only above this redshift \citep{vanderlinde10}, these systems are not used in the following analysis. Two systems (SPT-CL J0440-4744 and SPT-CL J0441-4502) are excluded from this analysis as they are detected in SPT-SZ regions that have been masked due to their proximity to point sources, which can compromise the SZE signal-to-noise measurement.  In addition, we exclude the three clusters highlighted in magenta: SPT-CL J0417-4748, SPT-CL J0456-5116 and SPT-CL J0502-6048. These systems are strongly masked in the DES-SVA1 data; based on the SZE position, the masks cover 40\% of the total cluster region. As a result, the associated optical counterparts are
highly mis-centered, and the corresponding richness is severely biased. We note that the average centering failure rate caused by the detection mask is 12\% (3 clusters out of 25), in comparison to the corresponding rate in the SDSS RM catalog, which is $\approx$1\% - 2\%. The difference reflects the fact that SDSS has a much larger contiguous area, while SVA1 has a more aggressive star mask. We expect this failure rate will decrease as the DES coverage increases, and object masking improves. Furthermore, improvements will be made to the RM algorithm to estimate the masked area not only at the putative centre of the cluster, but at all possible centres. In this way, clusters at high risk of mask-induced mis-centering will be properly removed from the sample.

The B15 catalog contains only SPT-SZ candidates with $\xi \ge 4.5$.
In this work we also apply the matching algorithm to SPT-SZ candidates
at $4 <\xi < 4.5$.  We identify 26 matches in this signal to noise
range. The resulting probabilities of chance associations of the
  $\xi>4$ sample are shown as empty histograms in
  Figure~\ref{fig:probability}. Similarly to the $\xi>4.5$ case, we
exclude 11 of these systems, which have estimated probabilities
$p$ of chance associations $p>16\%$. For the 15 matched
  systems, the expected number of false associations is also smaller
  than one. The remaining {\it cleaned} sample is shown as red points
on the left panel of Figure~\ref{fig:match}. The resulting total
number of SPT-SZ and RM associations at $\xi>4$ is 40.  This number is
in good agreement with the expectation ($\sim36$) obtained using the
number of SPT-candidates above $\xi >4$ in the DES-SVA1 region
explored by RM (88) and correcting it by the expected number of noise
fluctuations ($\sim45$) and the number of clusters above $z>0.8$
($\sim$7).  We find that two $\xi<4.5$ SPT-SZ candidates, SPT-CL
J0501-4717 and SPT-CL J0439-5611, have probabilities of random
associations larger than $5\%$, and therefore it is not clear whether
these low richness associations are correct (see
Figure~\ref{fig:match}).

The right panel of Figure \ref{fig:match} contains a comparison of the redshifts from the RM catalogs with the redshifts published in B15 ($z_\textrm{SPT}$) for the same clusters (obtained through dedicated optical/NIR followup by the SPT team or taken from the literature).  Clusters with spectroscopic
redshifts are highlighted in green.  We note that the redshift estimates are
not biased for the clusters affected by masking (magenta points).
For SPT-SZ candidates with $4<\xi< 4.5$ the SPT collaboration did not
complete followup optical imaging, and therefore we adopt the
redshifts of the RM optical cluster counterpart.

Table \ref{tab:catalog} contains all SPT candidates with RM counterparts used in this work.  For newly confirmed SPT-SZ clusters \footnotetext[2]{SPT-CL J0423-5506 was previously identified in Song et al. (2012) and Reichardt et al. (2012) at a redshift $z=0.21 \pm 0.04$ with a signal-to-noise $\xi = 4.51$. The associated redshift estimated with the same analysis presented in B15 is $z=0.25 \pm 0.036$.}, the associated $z_\textrm{SPT}$ redshift is not given. We caution that the masses for low-redshift clusters ($z < 0.25$) may be underestimated due to filtering that is done to remove the noise component associated with the primary CMB.

\begin{table*}
\caption{SPT-SZ cluster candidates with RM counterpart. We report the SPT-ID (1), right ascension (2) and declination (3), SPT peak detection significance $\xi$ (4), corresponding core radius (5), richness $\lambda $\, (6), associated redshift from the RM catalog (7) and SPT catalog (8), and SPT derived masses (9). Coordinates are J2000.}
\begin{tabular}{lccrcrccr}
\multicolumn{1}{c}{SPT ID} & R.A. & DEC & \multicolumn{1}{c}{$\xi$} & $\theta_c$ [arcmin] & \multicolumn{1}{c}{$\lambda$} & $z_\textrm{RM}$ & $z_\textrm{SPT}$ & $M_{500} [10^{14} h_{70}^{-1} M_\odot]$\\
\hline
SPT-CL J0438-5419 & $69.574$ & $-54.319$ & $22.88$ & $0.50$  & $144.76 \pm 5.52$ & $0.42 \pm 0.01$ & $0.42$ & $10.19 \pm 1.33$\\
SPT-CL J0040-4407 & $10.199$ & $-44.133$ & $19.34$ & $0.50$  & $137.45 \pm 7.03$ & $0.37 \pm 0.01$ & $0.35$ & $9.71 \pm 1.28$\\
SPT-CL J0417-4748 & $64.344$ & $-47.812$ & $14.24$ & $0.25$  & $54.22 \pm 6.75$ & $0.58 \pm 0.01$ & $0.58$ & $7.41 \pm 1.00$\\
SPT-CL J0516-5430 & $79.149$ & $-54.510$ & $12.41$ & $1.50$  & $178.93 \pm 8.71$ & $0.33 \pm 0.02$ & $0.29$ & $7.05 \pm 0.97$\\
SPT-CL J0449-4901 & $72.273$ & $-49.023$ & $8.91$ & $0.50$  & $91.37 \pm 4.75$ & $0.80 \pm 0.01$ & $0.79$ & $5.24 \pm 0.78$\\
SPT-CL J0456-5116 & $74.115$ & $-51.275$ & $8.58$ & $1.00$  & $73.08 \pm 5.39$ & $0.56 \pm 0.01$ & $0.56$ & $5.39 \pm 0.81$\\
SPT-CL J0441-4855 & $70.450$ & $-48.917$ & $8.56$ & $0.50$  & $86.96 \pm 4.55$ & $0.81 \pm 0.01$ & $0.79 \pm 0.04$ & $5.10 \pm 0.77$\\
SPT-CL J0439-4600 & $69.807$ & $-46.012$ & $8.28$ & $0.25$  & $55.18 \pm 3.52$ & $0.34 \pm 0.02$ & $0.34 \pm 0.04$ & $5.52 \pm 0.84$\\
SPT-CL J0440-4657 & $70.229$ & $-46.964$ & $7.13$ & $1.25$  & $67.95 \pm 3.62$ & $0.33 \pm 0.02$ & $0.35 \pm 0.04$ & $4.95 \pm 0.81$\\
SPT-CL J0447-5055 & $71.843$ & $-50.921$ & $5.97$ & $0.25$  & $77.84 \pm 5.26$ & $0.40 \pm 0.01$ & $0.39 \pm 0.05$ & $4.24 \pm 0.81$\\
SPT-CL J0422-5140 & $65.591$ & $-51.674$ & $5.86$ & $1.00$  & $49.28 \pm 5.32$ & $0.58 \pm 0.01$ & $0.59 \pm 0.03$ & $3.98 \pm 0.78$\\
SPT-CL J0439-5330 & $69.928$ & $-53.502$ & $5.61$ & $0.75$  & $60.77 \pm 3.81$ & $0.43 \pm 0.01$ & $0.43 \pm 0.04$ & $3.97 \pm 0.81$\\
SPT-CL J0433-5630 & $68.249$ & $-56.502$ & $5.32$ & $1.75$  & $60.75 \pm 4.82$ & $0.71 \pm 0.02$ & $0.69$ & $3.56 \pm 0.78$\\
SPT-CL J0535-5956 & $83.791$ & $-59.939$ & $5.20$ & $0.25$  & $50.25 \pm 4.09$ & $0.67 \pm 0.02$ & $0.58 \pm 0.03$ & $3.46 \pm 0.77$\\
SPT-CL J0440-4744 & $70.242$ & $-47.736$ & $5.12$ & $1.25$  & $82.55 \pm 3.80$ & $0.30 \pm 0.02$ & $-$ & $3.75 \pm 0.83$\\
SPT-CL J0428-6049 & $67.026$ & $-60.828$ & $5.11$ & $1.25$  & $55.91 \pm 5.95$ & $0.73 \pm 0.02$ & $0.64 \pm 0.03$ & $3.46 \pm 0.79$\\
SPT-CL J0444-4352 & $71.162$ & $-43.872$ & $5.01$ & $1.50$  & $70.53 \pm 5.91$ & $0.57 \pm 0.01$ & $0.57 \pm 0.03$ & $3.53 \pm 0.82$\\
SPT-CL J0458-5741 & $74.598$ & $-57.695$ & $4.87$ & $2.50$  & $37.90 \pm 2.68$ & $0.19 \pm 0.01$ & $0.19 \pm 0.02$ & $3.69 \pm 0.85$\\
SPT-CL J0534-5937 & $83.606$ & $-59.625$ & $4.74$ & $0.25$  & $40.43 \pm 3.42$ & $0.58 \pm 0.01$ & $0.58$ & $3.15 \pm 0.76$\\
SPT-CL J0502-6048 & $75.724$ & $-60.810$ & $4.69$ & $0.25$  & $30.73 \pm 4.32$ & $0.79 \pm 0.02$ & $-$ & $3.03 \pm 0.76$\\
SPT-CL J0441-4502 & $70.345$ & $-45.040$ & $4.62$ & $2.50$  & $51.22 \pm 4.11$ & $0.15 \pm 0.01$ & $-$ & $3.49 \pm 0.85$\\
SPT-CL J0429-5233 & $67.430$ & $-52.559$ & $4.56$ & $0.75$  & $33.84 \pm 3.97$ & $0.52 \pm 0.01$ & $0.53 \pm 0.03$ & $3.15 \pm 0.79$\\
SPT-CL J0452-4806 & $73.002$ & $-48.108$ & $4.52$ & $0.50$  & $56.54 \pm 4.89$ & $0.42 \pm 0.01$ & $0.37 \pm 0.04$ & $3.26 \pm 0.81$\\
SPT-CL J0437-5307 & $69.259$ & $-53.119$ & $4.51$ & $0.25$  & $36.89 \pm 3.56$ & $0.29 \pm 0.02$ & $-$ & $3.20 \pm 0.80$\\
SPT-CL J0500-4551 & $75.209$ & $-45.856$ & $4.51$ & $0.75$  & $32.68 \pm 4.82$ & $0.26 \pm 0.01$ & $-$ & $3.66 \pm 0.91$\\
SPT-CL J0453-5027 & $73.307$ & $-50.451$ & $4.47$ & $0.25$  & $31.99 \pm 3.44$ & $0.77 \pm 0.02$ & $-$ & $2.89 \pm 0.74$\\
SPT-CL J0449-4440 & $72.473$ & $-44.672$ & $4.37$ & $0.75$  & $54.50 \pm 5.43$ & $0.15 \pm 0.00$ & $-$ & $3.42 \pm 0.86$\\
SPT-CL J0423-5506\footnotemark[2] & $65.809$ & $-55.104$ & $4.36$ & $1.25$  & $38.65 \pm 3.44$ & $0.27 \pm 0.02$ & $-$ & $3.26 \pm 0.83$\\
SPT-CL J0451-5057 & $72.937$ & $-50.965$ & $4.34$ & $0.50$  & $83.11 \pm 4.62$ & $0.76 \pm 0.01$ & $-$ & $2.81 \pm 0.74$\\
SPT-CL J0438-4629 & $69.564$ & $-46.488$ & $4.31$ & $0.50$  & $41.01 \pm 3.37$ & $0.43 \pm 0.01$ & $-$ & $3.07 \pm 0.79$\\
SPT-CL J0456-4531 & $74.099$ & $-45.523$ & $4.30$ & $0.25$  & $32.45 \pm 3.18$ & $0.29 \pm 0.02$ & $-$ & $3.17 \pm 0.81$\\
SPT-CL J0431-5353 & $67.970$ & $-53.896$ & $4.22$ & $0.50$  & $57.23 \pm 4.76$ & $0.75 \pm 0.02$ & $-$ & $2.74 \pm 0.73$\\
SPT-CL J0501-4717 & $75.274$ & $-47.294$ & $4.20$ & $3.00$  & $19.47 \pm 2.50$ & $0.35 \pm 0.02$ & $-$ & $3.34 \pm 0.88$\\
SPT-CL J0518-5740 & $79.507$ & $-57.670$ & $4.19$ & $0.25$  & $65.22 \pm 4.82$ & $0.82 \pm 0.01$ & $-$ & $2.60 \pm 0.70$\\
SPT-CL J0438-4907 & $69.655$ & $-49.117$ & $4.19$ & $1.75$  & $76.20 \pm 4.16$ & $0.24 \pm 0.01$ & $-$ & $3.13 \pm 0.81$\\
SPT-CL J0513-5901 & $78.273$ & $-59.029$ & $4.17$ & $0.25$  & $33.29 \pm 3.58$ & $0.61 \pm 0.01$ & $-$ & $2.75 \pm 0.73$\\
SPT-CL J0451-4910 & $72.888$ & $-49.178$ & $4.14$ & $0.25$  & $54.10 \pm 4.10$ & $0.73 \pm 0.02$ & $-$ & $2.71 \pm 0.73$\\
SPT-CL J0439-5611 & $69.978$ & $-56.192$ & $4.14$ & $0.50$  & $17.82 \pm 2.93$ & $0.28 \pm 0.02$ & $-$ & $3.10 \pm 0.81$\\
SPT-CL J0532-5752 & $83.237$ & $-57.877$ & $4.11$ & $0.50$  & $48.49 \pm 3.81$ & $0.77 \pm 0.02$ & $-$ & $2.59 \pm 0.71$\\
SPT-CL J0449-5908 & $72.472$ & $-59.142$ & $4.11$ & $1.25$  & $107.14 \pm 5.29$ & $0.77 \pm 0.01$ & $-$ & $2.68 \pm 0.73$\\
\hline
\end{tabular}
\label{tab:catalog}
\end{table*}

\section{Mass Calibration Method}
\label{sec:method}
We apply the method described in Bocquet et al. (2015) to characterize the $\lambda$-mass relation of SPT-selected clusters. We refer the reader to the original paper for a detailed description of the method. A similar approach has been adopted by \citet{liu14} for studying the SZE properties of an X-ray selected cluster sample from the XMM-BCS survey \citep{suhada12,desai12}.  In this analysis, we consider the RM richness as a follow-up observable to the SZE-selected cluster sample.  This choice is adequate as there are no SPT-SZ candidates with $\xi>4.5$ missing RM counterparts in the redshift and spatial regime explored by the RM catalog, so that the cross-sample can indeed be thought of as solely SPT-selected. We note that this is not the case for SPT-SZ candidates with $4<\xi \le 4.5$ that do not have RM counterparts. However, the adopted method is also accurate under the assumption that cross-matching the SPT-SZ candidate list with the RM cluster catalog {\it cleans} the SPT-SZ candidate list, removing the expected noise fluctuations.  Within this context the resulting cluster sample is therefore drawn from the halo mass function through the SPT-SZ selection in the redshift range explored by the RM catalog.

In the following subsections we describe the model we use to simultaneously constrain the SZE-mass relation (Section \ref{sec:sze_mass}) and the richness-mass relation (Section \ref{sec:richness_mass}).

\subsection{The SZE-mass Relation}
\label{sec:sze_mass}
Following previous SPT papers
    \citep[][B15]{vanderlinde10,benson11,reichardt13,bocquet15}, we define
    the unbiased SZE significance $\zeta$ as the average
    signal-to-noise a cluster would produce over many realizations of
    SPT data, if the cluster position and core radius were perfectly
    known. This quantity is related to the expectation value of $\xi$
    over many realizations of the SPT data by: 
 \be \zeta=\sqrt{\langle \xi \rangle^2 -3},
\label{eq:xi}
\ee where the bias in $\langle \xi \rangle$ is due to
  maximizing the signal-to-noise over three variables (cluster right ascension,
  declination, and core radius). The scatter of the actual observable
  $\xi$ with respect to $\langle \xi \rangle$ is characterized by a
  Gaussian of unit width.
The SPT observable-mass relation
$P(\zeta|M_{500},z)$ is modeled as a log-normal distribution of mean
\begin{eqnarray}
 \langle \textrm{ln} \zeta| M_{500}, z \rangle&=& \textrm{ln}
A_\textrm{SZE} + {B_{\textrm{SZE}}}\, \textrm{ln} \left ( \frac{M_{500}}{3
  \times10^{14} h^{-1} \msun} \right ) \nonumber \\
  & & + {C_{\textrm{SZE}}}\, \textrm{ln}
\left ( \frac{E(z)}{E(z=0.6)} \right )
\label{eq:zeta}
\end{eqnarray}
and scatter $D_{\textrm{SZE}}$, and where $E(z)\equiv H(z)/H_0$.
At low significance $\zeta \lesssim 2$, there is a non-negligible chance of multiple low-mass clusters overlapping within the same resolution element of the SPT beam. We account for this by only considering the brightest of these objects per approximate resolution element and we compute $P( \zeta_\textrm{max}| \zeta) $
 following \cite{crawford10}. The SPT observable-mass relation is therefore expanded to $P( \zeta_\textrm{max}| \zeta)P(\zeta|M_{500},z)$ and $\zeta_\textrm{max}$ is then converted to the observable $\xi$ as in Eq.3.

To calibrate the $\zeta$--$M$ relation we use the subsample of clusters with $\xi>5$ and $z>0.25$ from the 2500d SPT-SZ catalog (B15).  We determine the parameter values by abundance-matching the catalog against our fixed reference cosmology.  We predict the expected number of clusters as a function of mass and redshift using the halo mass function \citep{tinker08}. We convolve this mass function with the observable-mass relation accounting for its associated uncertainties, and compare the prediction with the data.  Our approach here is effectively the opposite of the typical analysis, where cosmological parameters are deduced from the cluster sample using both priors and calibrating information to constrain the scaling relation parameters \citep[e.g.][]{benson11,bocquet15}; here, we assume perfect knowledge of cosmology to calibrate the scaling relation. Note that this method does not depend on any assumptions about hydrostatic equilibrium.

We assume flat priors on $A_\textrm{SZE}$, $B_\textrm{SZE}$, $C_\textrm{SZE}$ and a Gaussian prior on $D_\textrm{SZE} = 0.18\pm0.07$; the latter corresponds to the posterior distribution derived from the cosmological analysis of the full SPT sample (de Haan et al., in preparation).  We obtain the following parameters for the $\zeta$-mass relation by maximizing the likelihood of obtaining the observed sample in $\xi$ and redshift under the model derived from Eq. 3 and Eq. 4.  The results are
\begin{eqnarray}
&A_\textrm{SZE} = 4.02 \pm 0.16, B_\textrm{SZE}= 1.71 \pm 0.09,&\nonumber\\
&C_\textrm{SZE} = 0.49 \pm 0.16, D_\textrm{SZE}=0.20 \pm 0.07.&
\end{eqnarray}

For every cluster in the sample we also calculate the associated mass distribution, accounting for selection effects: 
\be
P(M_{500}|\xi,z , \vec p) \propto P(\xi|M_{500},z,\vec p) \, P(M_{500}|z,\vec p), 
\ee
where the vector $\vec p$ encapsulates cosmological and scaling relation parameters and $P(\xi|M_{500},z,\vec p)$ is obtained from the $\xi$-mass scaling relation as described by Eq. ~\ref{eq:xi} and \ref{eq:zeta}. The halo mass function $P(M_{500}|z,\vec p)$ is the prior on the mass distribution at redshift $z$.

Masses derived for the matched cluster sample are shown in Table~1. We note that both these masses and the SZE scaling relation parameters quoted here are different from the ones reported in B15. We adopt the same fixed cosmology as in B15, but in this analysis we consider data from the full SPT-SZ cluster-survey as opposed to just the sample from the initial 720~\degs \citep{reichardt13}. 

\subsection{Richness-mass Relation}
\label{sec:richness_mass}
As for the SZE-mass relation (Eq.~\ref{eq:zeta}), we assume a power law form for the \lam-mass relation: 
\begin{eqnarray}
\langle \textrm{ln} \lambda| M_{500}, z \rangle & = &  \textrm{ln} A_\lambda + B_\lambda\, \textrm{ln} \left ( \frac{M_{500}}{3\times10^{14} h^{-1}\msun} \right ) \nonumber \\
& & + C_\lambda\, \textrm{ln} \left ( \frac{E(z)}{E(z=0.6)} \right )
\label{eq:lambda_scaling}
\end{eqnarray}
where $A_\lambda$ is the normalization, $B_\lambda$ characterizes the mass dependence, and $C_\lambda$ characterizes the redshift evolution.  An additional parameter $D_\lambda$ describes the intrinsic scatter in $\lambda$, which is assumed to be log-normal and uncorrelated with the SZE scatter, with variance given by:
\be
\textrm{Var}(\textrm{ln} \lambda|M_{500}) = \textrm{exp} ( - \langle \textrm{ln} \lambda|M_{500} \rangle ) + D_\lambda^2.
\label{eq:lambda_scatter}
\ee 
The first term above represents the Poisson noise associated with the number of galaxies in a halo at fixed mass, and therefore we define the intrinsic scatter $D_\lambda$ as log-normal scatter in addition to Poisson noise.  We assume flat priors
on the distributions of $A_\lambda$, $B_\lambda$, $C_\lambda$ and a positive flat prior for $D_\lambda$.

The probability that a cluster with SPT-SZ signal-to-noise $\xi$ is
observed to have a richness $\lambda$ is \be P(\lambda|\xi,z,\vec p)
=\int{dM_{500}P(\lambda|M_{500},z,\vec p)P(M_{500}|\xi,z,\vec p)}. \ee
The term $P(\lambda|M_{500},z,\vec p)$ contains the lognormal
intrinsic scatter and normal measurement uncertainties in the
observable \lam.  We use the above distribution to evaluate the
likelihood of the matched cluster sample defined through our cross-matching
procedure.  Note that we simultaneously vary both the optical and SZE
scaling relation parameters, further including the SZE data set
from B15 with $\xi>5$, $z>0.25$ for constraining the
SZE--mass relation.

\begin{table}
\centering
\caption{Best fitting parameters and $68\%$ confidence level of the richness-mass scaling relation parameters described by Equation~\ref{eq:lambda_scaling} and \ref{eq:lambda_scatter}. }
\begin{tabular}{lcccc}
Catalog & $A_\lambda$ & $B_\lambda$ & $C_\lambda$ & $D_\lambda$\\
\hline\\[-7pt]
SPT-SZ+RM $\xi>4.5$ 	& $66.1_{-5.9}^{+6.3}$ 	& $1.14_{-0.18}^{+0.21}$ 	& $0.73_{-0.75}^{+0.77}$ 	& $0.15_{-0.07}^{+0.10}$ \\[3pt]
SPT-SZ+RM $\xi>4$ 		& $69.8_{-4.9}^{+6.0}$ 	& $1.17_{-0.17}^{+0.19}$ 	& $1.71_{-0.57}^{+0.63}$ 	& $0.20_{-0.08}^{+0.09}$  \\[3pt]
\hline
\end{tabular}
\label{tab:scaling}
\end{table}

\section{Results}
\label{sec:results}

We present here the constraints on the richness-mass relation (Section~\ref{sec:scaling-rel}) and then use these best fit parameters to explore whether the cumulative distribution of the matched samples are consistent with the expectations from the model (Section~\ref{sec:goodness}).  Finally, in Section~\ref{sec:offset} we analysis the optical-SZE positional offset distribution.

\subsection{redMaPPer Richness-mass Relation}
\label{sec:scaling-rel}

We marginalize over the SZE-mass scaling relation parameters and constrain the posterior distributions for the RM \lam-mass scaling relation.  Our best fit parameters and $68\%$ confidence level intervals are reported in Table {\ref{tab:scaling} and shown in Figure~\ref{fig:scaling_spt}. 
We note that the slope of the \lam-mass relation is consistent with 1 within  $1\sigma$ (consistent, therefore, with the richness being proportional to the mass), and the model is consistent with no redshift evolution within $1\sigma$ \citep{andreon14}. Furthermore the resulting \lam-mass relation is characterized by a remarkably low asymptotic intrinsic scatter, with $\sigma_{\textrm{ln}\lambda|M_{500}}\xrightarrow{} 0.15^{+0.10}_{-0.07}$ as $\langle \textrm{ln} \lambda| M_{500} \rangle \to\infty$ (Eq. \ref{eq:lambda_scatter}).
Following Evrard et al. (2014), we estimate the characteristic scatter in mass at fixed richness to arrive at
$\sigma_{\ln M} = 0.18^{+0.08}_{-0.05}$ at $\lam=70$, $\sim 25 \%$ larger than the corresponding characteristic scatter in mass at fixed $\xi$.
We present in Figure~\ref{fig:match_mass} the RM richness as a function of the SPT derived masses. 
Blue lines describe the best fitting model and intrinsic scatter as derived from this analysis (Table 2) at a pivot point of $z=0.6$.  
 
We have verified that our results are not dominated by uncertainties in the SZE-mass scaling relation by fixing these parameters to their best fit values. Our results are only marginally improved in this case. Consequently, future analyses with larger samples are expected to considerably reduce the uncertainties of the recovered \lam-mass scaling relation parameters.

\begin{figure}
\hbox to \hsize{
    \includegraphics[width=85mm]{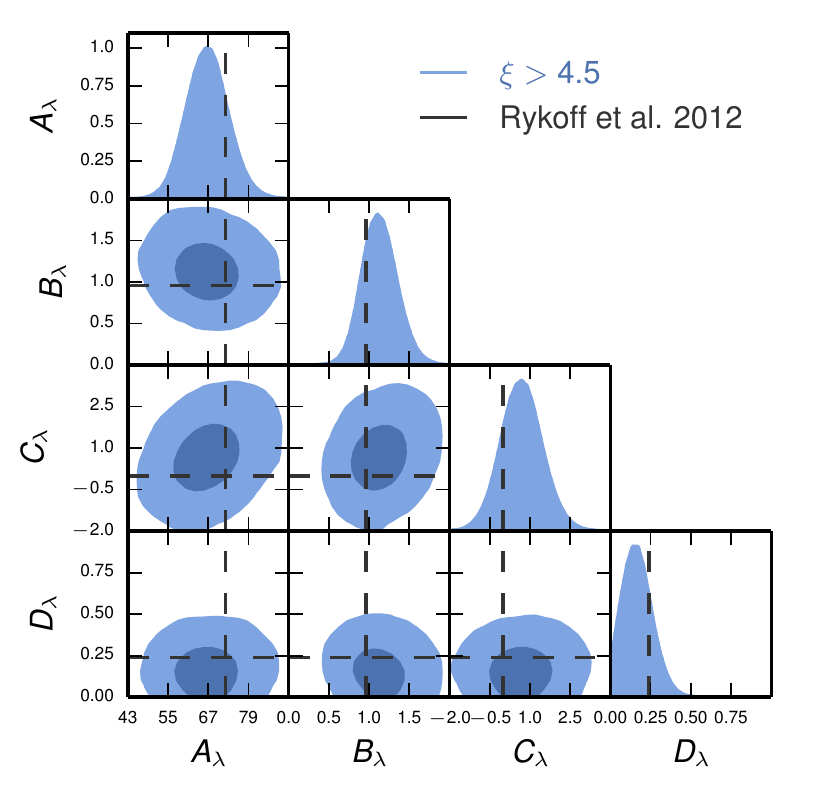}\hfil
 }
\caption{Posterior distribution for the four parameters of the \lam-mass scaling relation (Equations~\ref{eq:lambda_scaling} and \ref{eq:lambda_scatter}). Predictions from the scaling relation of Rykoff et al. (2012) are shown as dashed black lines.  Best fitting parameters and associated $1 \sigma$ uncertainties appear in Table~\ref{tab:scaling}.
\label{fig:scaling_spt}}
\end{figure}

In Figure~\ref{fig:match_mass}, one cluster appears to be an obvious outlier: SPT-CL J0516-5435 ($\xi=12.4, \lambda=178.9$,\mvir=7.05$\times 10^{14}\msunh$). 
SPT-CL J0516-5345 is a well-known merger that is elongated in a north-south direction in the plane of the sky with an X-ray mass estimate nearly a factor of two larger than the SZE mass estimate. This cluster was in fact the strongest outlier in the sample of 14 clusters in \citet{andersson11}. \cite{high12} made a weak lensing measurement of SPT-CL J0516-5345, and found that there was a significant offset between the brightest central galaxy (BCG) and the weak-lensing centre, consistent with the merger hypothesis.  Additionally, \cite{high12} found that the weak-lensing mass was in better agreement with the SZE-mass estimate than the X-ray mass estimate, at a level consistent with elongation observed in the plane of the sky. Therefore, SPT-CL J0516-5345 appears to be an outlier due to true intrinsic scatter in the observable-mass relations, so we leave it in our analysis. We note, however, that whether or not we include SPT-CL J0516-5345 in the fit has a significant impact on our results. Our best fit parameters shift from $B_\lambda = 1.14_{-0.18}^{+0.21}$ and $D_\lambda = 0.15_{-0.07}^{+0.10}$ with this cluster, to $B_\lambda = 1.00_{-0.15}^{+0.17}$ and $D_\lambda=0.05^{+0.07}_{-0.03}$ when SPT-CL J0516-5345 is not included in the fit.  Whether SPT-CL J0516-5345 represents a rare event in a non-Gaussian tail in the distribution of richness of galaxy clusters or the recovered log-normal scatter obtained when cluster SPT-CL J0516-5349 is included is more correct will thus need to await future analyses with larger samples.

\begin{figure}
\hbox to \hsize{
    \includegraphics[width=85mm]{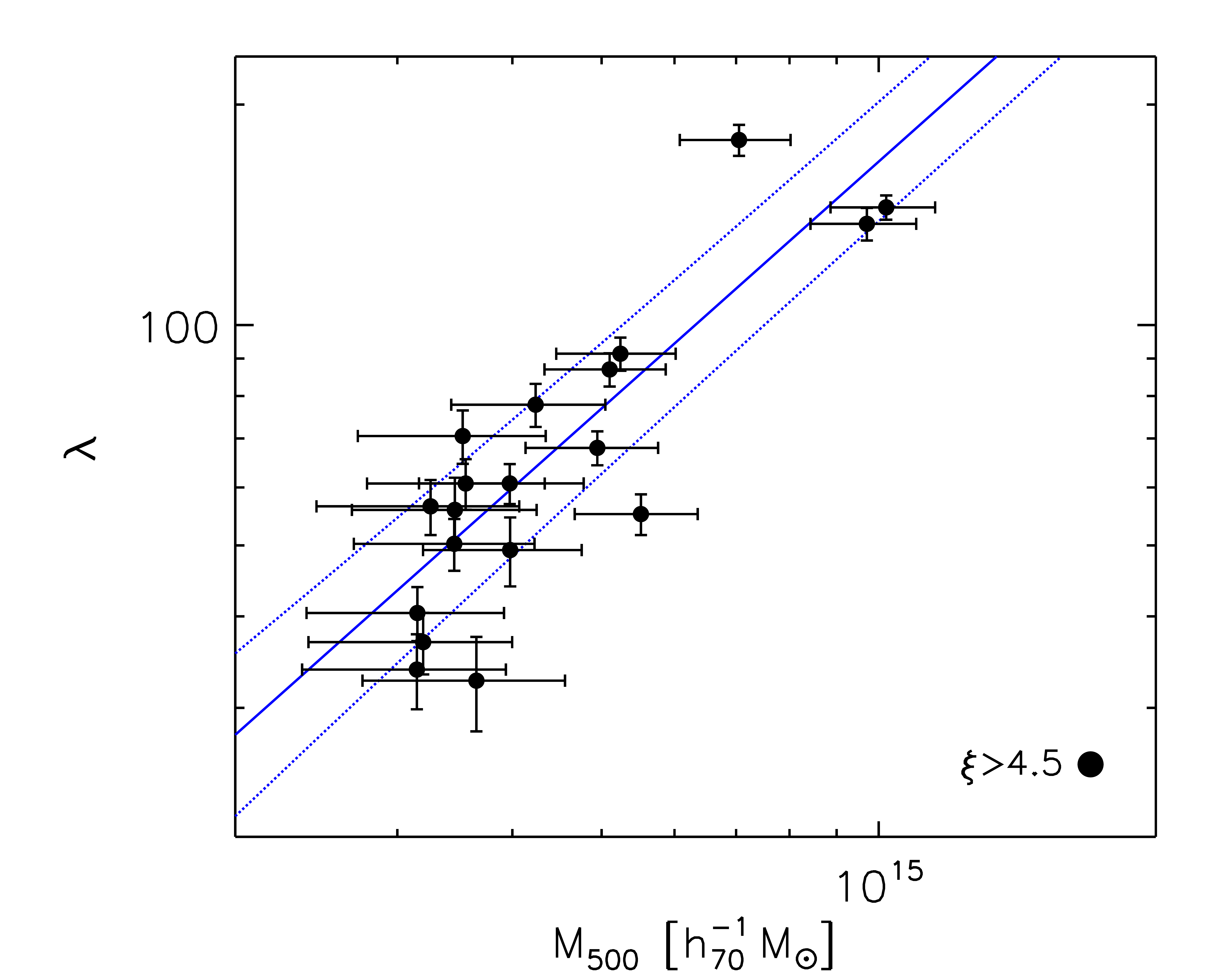}\hfil} 
    \vskip-0.1cm
\caption{Richness as a function of the SPT derived masses for the calibration sample used in this analysis (Section 2.3). Blue lines
  show the best fit richness-mass relation and $1 \sigma$ intrinsic
  scatter.} 
\label{fig:match_mass}
\end{figure}

We convert the $P(M_{500} | \lam)$ scaling relation derived by \citet{rykoff12} using abundance matching and the SDSS maxBCG cluster-catalog \citep{koester07a} to a richness--mass relation so that we can compare it to our results \citep[see also][]{evrard14}.  The predictions from \citet{rykoff12} are shown as dashed lines in Figure~\ref{fig:scaling_spt} under the assumption of no redshift evolution in the richness--mass relation.  We note that all parameters of our derived RM \lam-mass scaling relation for SPT-selected clusters are consistent with the \citet{rykoff12} values.

We repeat these analyses extending the sample to include those with
$4<\xi<4.5$ and find similar results (Table ~\ref{tab:scaling}).  The largest difference is
in the redshift evolution term which now has a best fit value $C_\lambda=1.71_{-0.57}^{+0.63}$. While formally this difference does not have large statistical significance ($1.3\sigma$), it is coming from a sample that includes a large fraction of the same clusters, so it is likely statistically significant.  A larger redshift evolution term would imply that higher redshift RM clusters are less massive at fixed \lam.  At the same time the derived scatter is also larger. 

Two of the clusters in the $4<\xi<4.5$ range are compatible with false associations.  Excluding the two matched clusters with the highest probability of random associations results in a $\sim 1\sigma$ shift in $B_\lambda$ (from $B_\lambda=1.17$ to $B_\lambda=1.04$) and in a $\sim 0.5 \sigma$ shift in $C_\lambda$ (from $C_\lambda=1.71$ to $C_\lambda=1.42$), while the other parameters ($A_\lambda$ and $D_\lambda$) are almost unchanged.  Interestingly, even though we have increased the number of clusters in the SPT-SZ+RM sample by $40\%$, the constraints on the scaling relation parameters are only mildly tighter. There are two reasons for this.  First these lower signal-to-noise SPT-SZ clusters have larger fractional mass uncertainties in $\xi$ ($\langle \xi \rangle ^{-1} \sim 0.23 $ and $\langle \xi \rangle ^{-1} \sim 0.13 $ respectively for the $4<\xi<4.5$ and $\xi>4.5$ samples). Second, the richnesses are also systematically lower, leading to a larger Poisson variance.  Thus, each low $\xi$ cluster has less constraining power than a high $\xi$ cluster, reducing the impact of extending the sample to include the lower mass systems.

\begin{figure*}
\hbox to \hsize{
    \includegraphics[width=180mm]{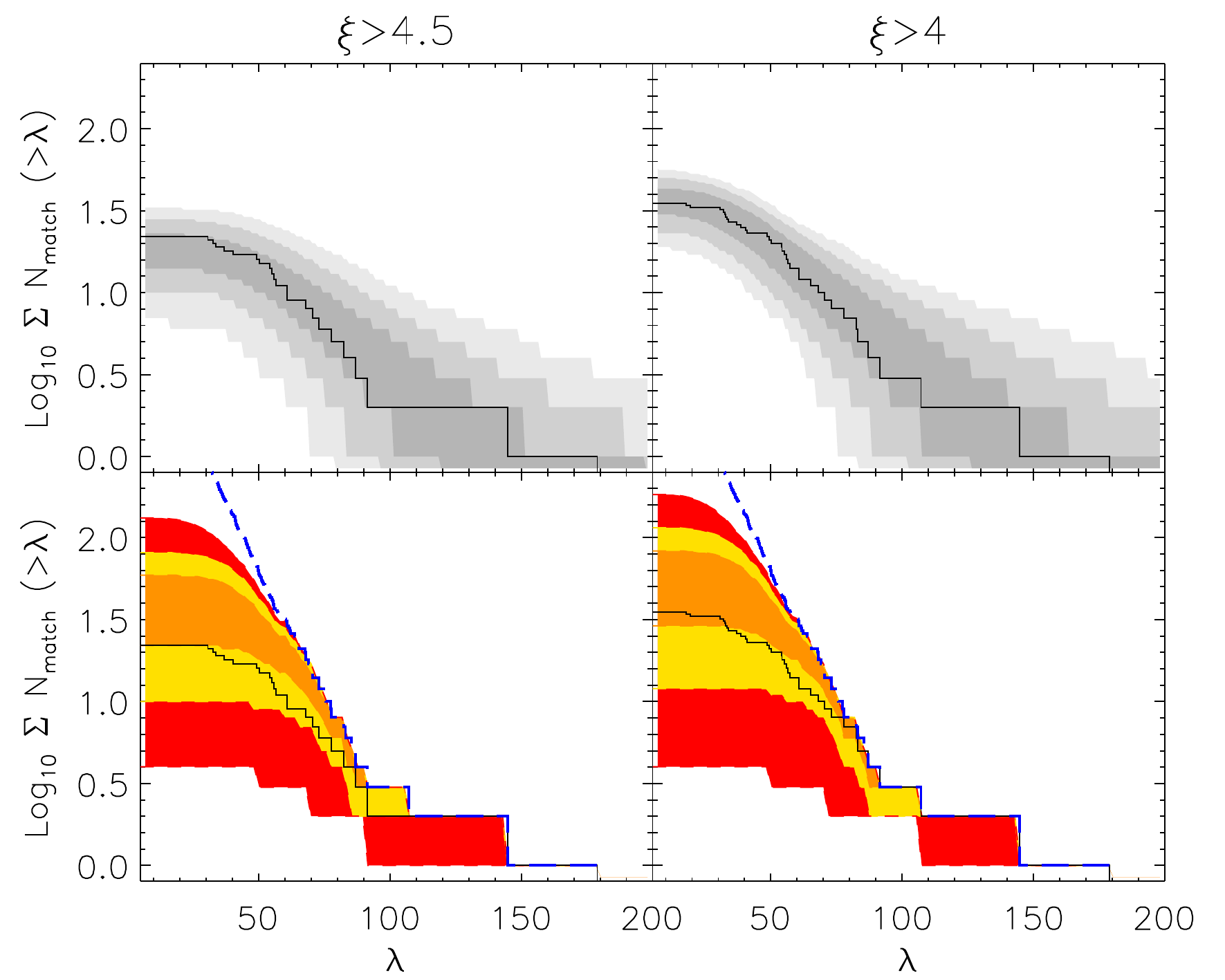}\hfil
     }
\caption{Consistency tests of our model (Section~\ref{sec:method}) and the SPT-SZ+RM catalog from the $\sim$124~\degs SPT-E field.  The solid black lines show the observed cumulative distribution in richness \lam\ of the $\xi>4.5$ (left) and $\xi>4$ (right) SZE-selected samples.  \textit{Upper panels:} Gray scale regions show 1, 2, and $3\sigma$ regions predicted by drawing 10$^6$ SZE-selected samples from the mass function and assigning \lam\ according to our scaling relation constraints.  There is good agreement with the data.  
\textit{Lower panels:} Dashed blue lines show the cumulative distribution in \lam\ of the full RM sample. Orange, yellow, and red areas define 1, 2, and $3\sigma$ regions representing the predicted cumulative distribution of the SPT-SZ+RM catalog using as input the full RM sample and the probability (Eq. \ref{eq:match}) that each RM cluster will have an SPT-SZ counterpart, given our scaling relation constraints. 

\label{fig:pred_match}}
\end{figure*}

\subsection{Consistency Test of Model}
\label{sec:goodness}
We also test the consistency of the adopted scaling relation model with the data by examining whether we are finding the expected number of matches with the correct distribution in richness.  To do this, we focus on the SPT-E field, which at $\sim 124.6$~\degs\ is the largest contiguous region covered by the DES-SVA1 data.  We carry out two different tests.

In the first test, we examine whether we are finding the expected number of SZE-selected clusters and whether these clusters have the expected number of optical matches with the correct \lam\ distribution.
We generate 10$^6$ Monte Carlo realizations of cluster samples extracted from the halo mass function above \mvir$>10^{13.5} \msunh$ and assign richnesses \lam\ and SPT-SZ significance $\xi$ using the parameters we extract from our analysis of the real matched catalog.  These Monte Carlo mocks are generated taking into account the survey area as a function of redshift sampled by RM in the SPT-E field.  We then apply the SZE selection--either $\xi>4.5$ or $\xi>4$--and measure the cumulative distribution in \lam\ of the SZE-selected samples.  We then compare this to the same distribution in the real matched catalog.  

Shaded regions in the upper panels of Figure~\ref{fig:pred_match} show 1, 2, and 3$\sigma$ confidence regions obtained from the mocks after marginalizing over the scaling relation parameters.  The solid black line shows the distribution from the real catalog, which is in good agreement with the mocks.  The largest observed difference is smaller than $1\sigma$ indicating that the adopted model provides a consistent description of the observed number and richness distribution of the SZE-selected sample. Similarly, a Kolmogorov-Smirnov (KS) test for the observed cumulative distribution of matched systems as a function of \lam \, and the corresponding median distribution from the Monte Carlo simulations, shows that the null hypothesis of data being drawn from same distributions cannot be excluded and returns $p-$values 0.76 and 0.96 for $\xi > 4.5$ and $\xi>4$, respectively. 

The second test is focused on whether the RM cluster catalog (which is significantly larger than the SPT-SZ catalog) has the expected number of SZE matches with the correct \lam\ distribution.  Essentially, we take the  observed RM catalog as a starting point, and calculate the expected number of systems with SPT-SZ counterparts given the model and parameter constraints from our \lam-mass likelihood analysis.  This test differs from the first in that it takes the \textit{observed} RM selected cluster sample as a starting point, and such a test should be more sensitive to, for example, contamination in the RM catalog. The formally correct way to evaluate a statistical difference between the RM selected sample and the SPT-SZ+RM matched sample used in this analysis would be to calibrate the \lam-mass and SZE-mass relations starting from the RM selected sample. Such a study goes, however, beyond the scope of the current work and will be addressed in a future project. Therefore, the following analysis is only intended to be a consistency check.

For this purpose, we proceed by first computing, for each real RM selected cluster in the SPT-E field, the probability $P_\textrm{m}$ of that cluster also having $\xi> 4.5$ and therefore being in the matched sample.  We define this probability as:
\begin{eqnarray} 
P_\textrm{m} & = & \int_{4.5}^{\infty}{P(\xi|\lambda,z) d\xi} \nonumber\\
& = & \int_{4.5}^{\infty}d\xi \int dM_{500} \, P(\xi|M_{500})P(M_{500}|\lambda,z),
\label{eq:match}
\end{eqnarray}
where $P(M_{500}|\lambda,z) \propto P(\lambda|M_{500},z)P(M_{500},z)$ and $P(M_{500},z)$ is the halo mass function \citep{tinker08}.  

To predict the expected number of RM clusters with SZE counterparts, we then randomly sample the scaling relation parameters, determining $P_\textrm{m}$ for all the RM clusters in each case. We then Monte Carlo sample each $P_\textrm{m}$ to produce randomly sampled matched cluster catalogs. We use the results from the ensemble of random matched catalogs to produce the expected cumulative distributions in \lam.  Orange, yellow, and red regions in the bottom panels of in Figure~\ref{fig:pred_match} show the 1, 2, and 3$\sigma$ confidence regions, respectively, of the expected cumulative distribution in \lam\ for $\xi>4.5$ (left) and $\xi>4$ (right), given the RM catalog and scaling relation parameter constraints as input.  The dashed blue line shows the cumulative distribution of the entire sample of RM clusters in the area, while the solid black line shows the observed cumulative distribution of the real matched catalog. 

We note that the predicted number of SPT-SZ+RM matches in this case tends to be higher than that observed for $\lam>35$, but the tension is weak. Of some concern is the high \lam\ end of the sample ($\lam > 70$), where only 9 of the 17 RM selected clusters have SPT-SZ counterparts at $\xi>4.5$ despite their having large probabilities indicating they should be in the SZE-selected sample. However, a KS test for the observed cumulative distribution of matched systems as a function of \lam \, and the corresponding median distribution from the mocks shows that the null hypothesis of data being drawn from the same parent distribution cannot be excluded and returns $p-$values 0.90 and 0.33, respectively, for $\xi > 4.5$ and $\xi>4$.

While the KS test is showing no evidence for the two distributions to differ, it is also known to be not very sensitive to the tails of the distributions \citep{2013arXiv1311.3190M}. With this respect, it is also therefore interesting to directly compare the observed cumulative distribution with the cumulative distribution from the predicted number of matches. In this case, as for the classical KS test, we focus on the largest difference between the two distributions. Lower left panel of Figure \ref{fig:pred_match} shows that at most, the observed cumulative distribution is in tension with expectations at the $\sim 2\sigma$ level, providing some indication of tension between the observed sample and our model. For example, when restricting ourselves to clusters with $\lambda>70$, we find that our simulation results in a larger number of SPT+RM associations in 94.6\% of our Monte Carlo realizations. This tendency to observe fewer matches than expected given the size of the RM selected sample could be explained by either contamination within the RM sample, additional incompleteness within the SPT-SZ sample beyond that caused by scatter in the SZE-mass relation, or simply a statistical fluctuation.  Future work exploring the SZE properties of the lower mass systems along with an extension of the current analysis to the full overlap between DES and the SPT-SZ survey will sharpen this test.

\subsection{Optical-SZE Positional Offset Distribution}
\label{sec:offset}
It has been shown that optical-miscentering can have a significant impact on the derived SZE signature for an optically-selected sample \citep{biesiadzinski12,sehgal13,rozo14b,rozo14c,rozo14d}.  We note, however, that in our case the SZE signal has been estimated at the SZE-determined position, so our results are not affected by optical miscentering.  In fact, we can now use our data to constrain the distribution of offsets between the SZE-determined and the optically-determined cluster centres.

As an example, we show in Figure~\ref{fig:cluster} the DES-SVA1 {\it
  gri} pseudo colour image of SPT-CL J0433-5630, an SPT-SZ selected cluster
with $\xi \sim 5.3$ at redshift $z = 0.69$ (B15). Yellow contours show the SPT-SZ signal-to-noise in steps of $\Delta \xi = 1$, while the magenta circle describes the
projected radius \rvir$/2$. The cyan label refers to the associated RM
cluster centre.  This RM cluster has richness $\lam \sim 60$. We note that the
most probable central galaxy selected by RM is significantly offset
from the SZE defined centre. As a result, the measured SZE signature
at the optical position ($\xi = 4.1$) would be significantly underestimated with respect to the derived unbiased quantity $\zeta = 5$ obtained through Eq. 3. We
stress that this effect is not important for the scaling relation results reported in Section \ref{sec:scaling-rel}, as the sample analyzed here is SZE-selected.

\begin{figure}
\centering
\includegraphics[width=83mm]{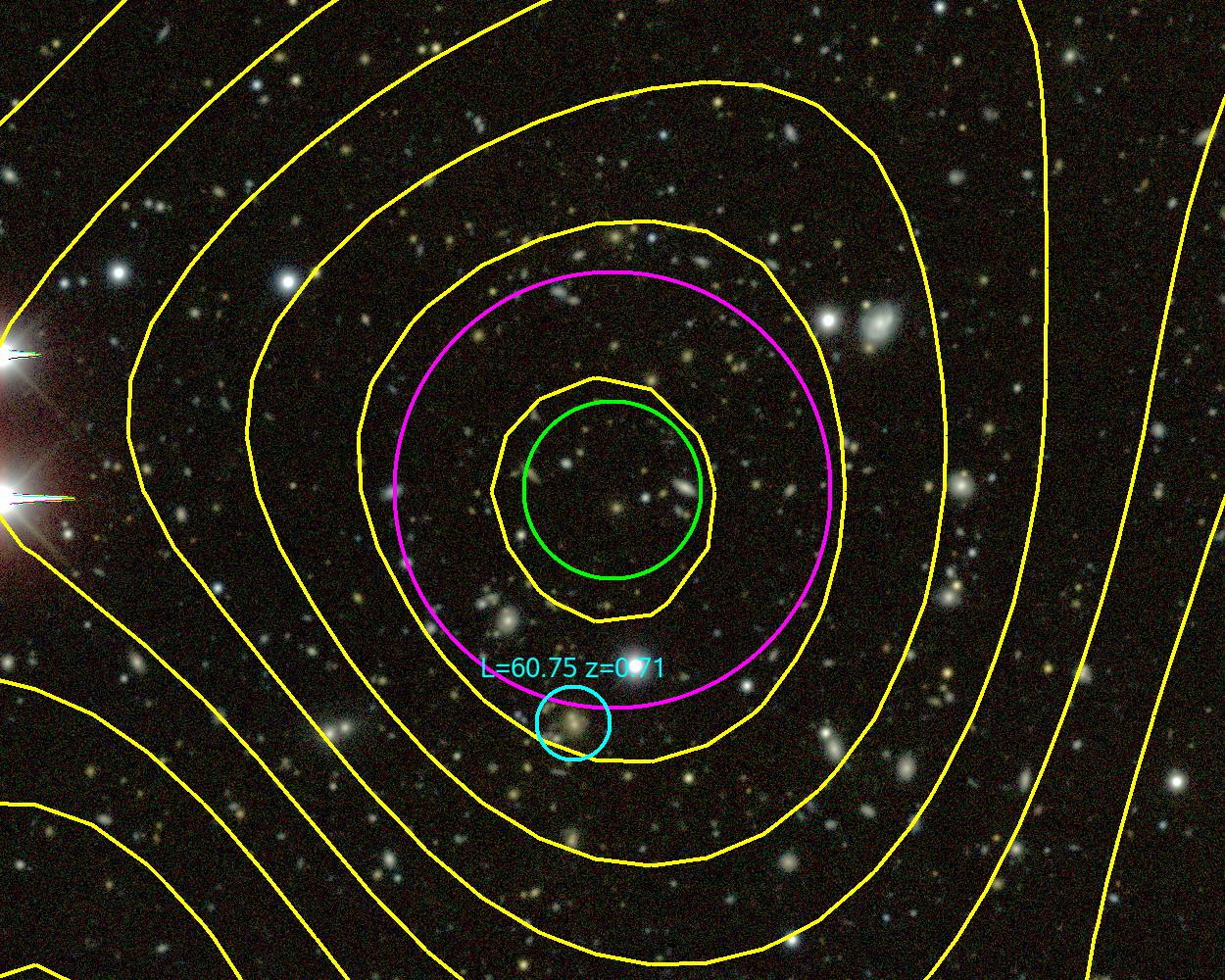}
\caption{SPT-CL J0433-5630: DES-SVA1 {\it gri} pseudo colour image overplotted with SPT-SZ signal-to-noise contours (in steps $\Delta \xi$=1). The magenta circle shows the projected \rvir$/2$ radius at $z=0.69$, while the green circle describes the $1\sigma$ SPT positional uncertainty (Eq.\ref{eq:uncert}). The cyan label marks the associated $\lam\sim60$ RM cluster.
\label{fig:cluster}}
\end{figure}

Figure~\ref{fig:histo} contains a normalized histogram of the distribution
of cluster positional offsets in units of \rvir\ for the $\xi>4.5$
analyzed SPT-SZ sample. Under the assumption that the
measurement uncertainty from the optical side is negligible, we model
this distribution as an underlying intrinsic positional offset
distribution convolved with the SPT-SZ positional uncertainty. 

The $1\sigma$ SPT-SZ positional uncertainty for a cluster with a pressure
profile given by a spherical $\beta$ model with $\beta = 1$ and core
radius $\theta_c$, detected with SPT-SZ significance $\xi$ is described
by: \be \Delta \theta = \xi^{-1}\sqrt{\theta_\textrm{beam}^2 +
  \theta_c^2 },
\label{eq:uncert}
\ee where $\theta_\textrm{beam}=1.2$ arcmin is the beam FWHM (see
\citealt{story11} and \citealt{song12} for more details).  As a
result, the expected distribution of positional offsets in the case in
which the intrinsic one is a $\delta-$function is shown (arbitrarily
rescaled) as a green line.

\begin{figure}
\centering
\includegraphics[width=87mm]{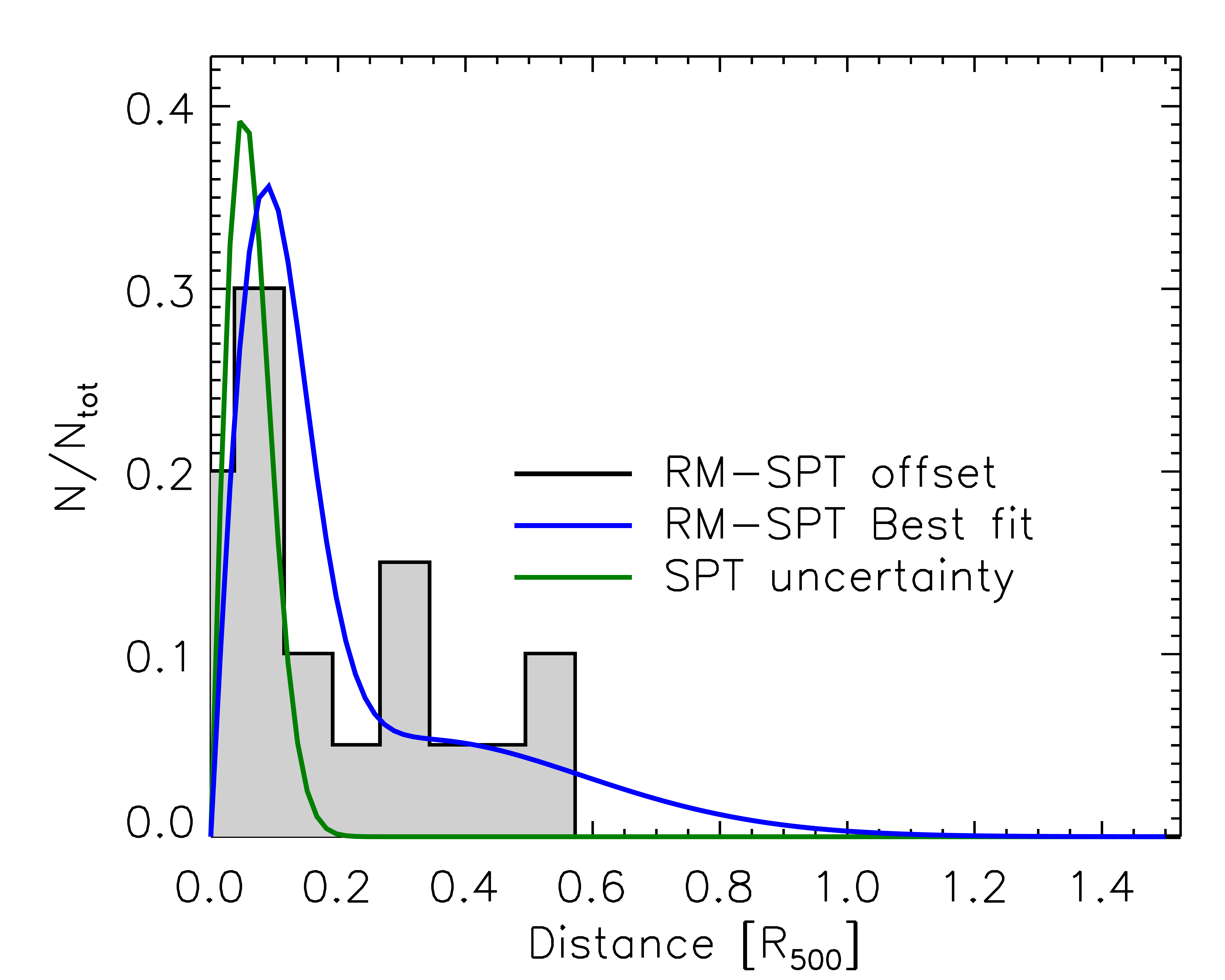}
\caption{Solid histogram shows the measured fraction of SPT-SZ+RM clusters as a function of the optical-SZE positional offset in units of \rvir. The green curve shows the SPT-SZ positional uncertainty, and the blue curves shows the best fitting SZE-optical positional offset model.
\label{fig:histo}}
\end{figure}

\citet{song12} have shown that the intrinsic optical-SZE positional offset distribution for an SPT-SZ selected sample is consistent with the optical$-$X-ray positional offset distribution of X-ray selected clusters \citep{lin04a}. 
The offset distribution can be characterized by a large population of central galaxies with small offsets from the SZE centres and a less populated tail of central galaxies with large offsets \citep[e.g.][]{lin04a,rozo14b,lauer14}.  We therefore parametrize the distribution of positional offsets between the RM centre and the SZE centre for $x$ as:
\be
P(x) = 2\pi x \left( \frac{\rho_0}{2 \pi \sigma_0^2}e^{{-\frac{x^2}{2 \sigma_0^2}} } + \frac{1- \rho_0}{2 \pi \sigma_1^2}e^{{-\frac{x^2}{2 \sigma_1^2}} } \right) 
\label{eq:modoff}
\ee where $x = r/$\rvir. While this model for the distribution was motivated by the expected intrinsic positional offset distribution, the measured distribution will include both the actual physical SZE-central galaxy offset distribution and the systematics due to failures in identifying the correct cluster center with the RM algorithm. For every cluster and parameter $\rho_0 \in [0,1]$, $\sigma_0 \in [0,1]$, and $\sigma_1 \in [\sigma_0,1]$, we then compare the predicted offset distribution obtained by convolving the model with the SPT-SZ positional uncertainty of Eq.~\ref{eq:uncert} to extract the associated likelihood.  Best fit parameters and $68\%$ confidence intervals are shown in Table~\ref{tab:off} and  joint and fully marginalized parameter constraints are shown in Figure~\ref{fig:offce}.

\begin{table}
\caption{Best fitting parameters and $68\%$ confidence level of the optical-SZE positional offset distribution.}
\begin{tabular}{lllll}
Catalog & $\rho_0$ & $\sigma_0 [$\rvir] & $\sigma_1 [$\rvir]& \\
\hline\\[-7pt]
RM-$\xi>4.5$ 	& $0.63^{+0.15}_{-0.25}$ 	& $0.07^{+0.03}_{-0.02}$ & $0.25^{+0.07}_{-0.06}$\\[3pt]
\hline
\end{tabular}
\label{tab:off}
\end{table}

\begin{figure}
\centering
\includegraphics[width=85mm]{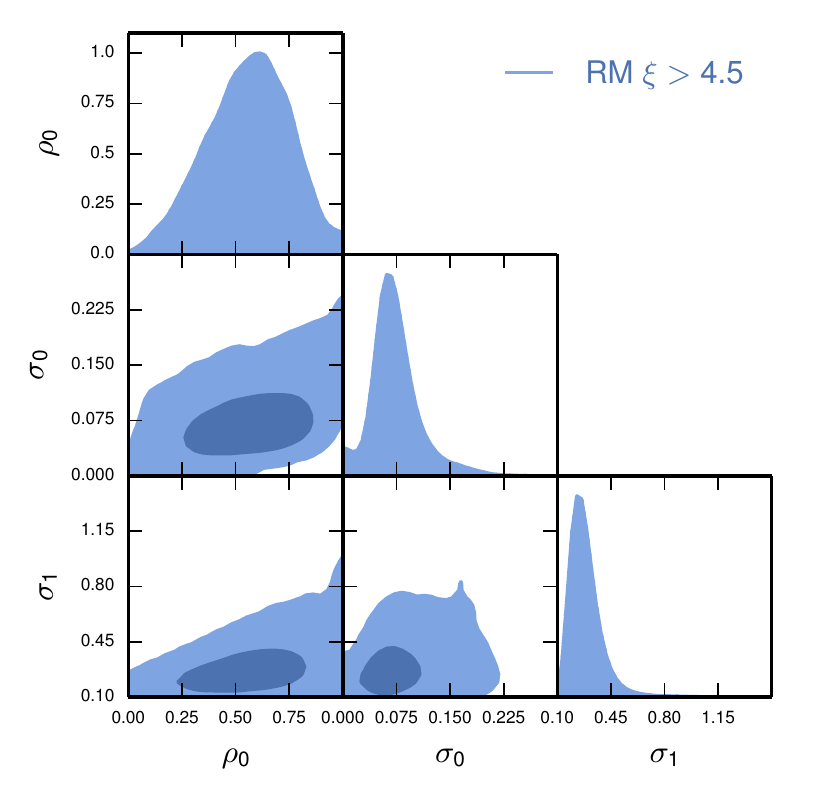}
\caption{Posterior distribution for the 1 and 2$\sigma$ level of the three parameter model describing the positional offset distribution of Equation~\ref{eq:modoff}. Best fitting parameters are shown in Table \ref{tab:off}. 
\label{fig:offce}}
\end{figure}

We note that the positional offset distribution for the RM sample is consistent with a concentrated dominant population ($\rho_0 =0.63^{+0.15}_{-0.25}$) of smaller offsets ($\sigma_0 = 0.07^{+0.03}_{-0.02}$ \rvir) and a sub-dominant population characterized by larger offsets ($\sigma_1 = 0.25^{+0.07}_{-0.06}$ \rvir). In the limit where the SZ center and BCG are coincident for every cluster, the RM code\citep{rykoff12,rykoff14} predicts the SZ-RM offset distribution for this sample to contain a centrally peaked component with normalization $\langle P_\textrm{cen} \rangle$ to be 0.79. While this is formally consistant with the measured SZ - RM offset distribution, intrinsic optical-SZE positional offsets, such as those presented by \cite[e.g.,][]{song12} likely contribute to large separation component of the observed distribution shown in Figure~\ref{fig:histo}. Larger samples will be necessary to disentangle the impact of systematics due to failures in identifying the correct cluster centre with the RM algorithm from the offset distribution due to cluster morphology.

\section{Conclusions}
\label{sec:conclusions}
In this paper, we cross-match SZE-selected cluster candidates with $\xi>4.5$ from the 2500d SPT-SZ survey (B15) with the optically-selected cluster catalog extracted from the DES science verification data DES-SVA1.  The optically-selected catalog is created using the RM cluster-finding algorithm. 

We study the robustness of our matching algorithm by applying it to randomly generated RM catalogs.

Using the adopted matching algorithm in the 129.1~\degs\ of overlap
between the two data sets, we create a matched catalog of 33 clusters.
Eight of these clusters are removed as likely chance superpositions
that are identified using the randomly generated catalogs.  The
resulting 25 cluster sample includes all previously known $z<0.8$ and
$\xi>4.5$ SPT-SZ clusters in this area \citep[][B15]{song12,reichardt13} in addition to three previously unconfirmed SPT-SZ clusters.

We then study three characteristics of this cross-matched SPT-SZ+RM cluster sample:

1) \textit{The richness mass relation of SPT-SZ selected clusters.}
We calibrate the \lam-mass relation from SZE measurements by
  applying the method described in Bocquet et al. (2015). In this
analysis we assume a fixed fiducial cosmology and marginalize over the
simultaneously calibrated SPT-SZ $\xi$-mass relation. We
adopt flat priors on the richness--mass relation parameters. 
We find that the RM \lam-mass relation for SPT-SZ selected clusters is
characterized by a small asymptotic intrinsic scatter
$D_\lambda=0.15_{-0.07}^{+0.10}$ and by a slope $B_\lambda=1.14_{-0.18}^{+0.21}$
that is consistent with unity. Our constraints are in good agreement with
those of \citet{rykoff12} and show that the scatter in mass at fixed
richness $\lam=70$ for this sample is only $25\%$ larger than the scatter
in mass at fixed SPT-SZ observable $\xi$.

2) \textit{Consistency test of model and matched catalog.}  We carry out two consistency tests to determine whether there is tension between the observed matched sample and the expectations given the scaling relation model we have adopted.  Both tests involve creating Monte Carlo generated
cluster catalogs with associated richness and SPT-SZ significance derived from
the fitted scaling relations. The first test checks whether the correct number of SZE-selected clusters is found and whether those clusters exhibit the correct number of optical matches with the expected \lam\ distribution. As is clear from Figure~\ref{fig:pred_match}, the observations are perfectly consistent with the expectations from the model. Thus, our analysis shows that the data in our matched SPT-SZ+RM sample are well described by our adopted model. In the second test we take the much larger \textit{observed} RM catalog as a starting point and use the model to test whether the expected number of SZE matches with the expected \lam\ distribution is found.  Unlike the first test, this one would in principle be sensitive to contamination within the RM sample.  Here the agreement is not as good because there is a tendency for there to be fewer observed matches than expected.  However, the tension reaches the $\sim 2\sigma$ level at worst, and so there is no convincing evidence that our observed sample is inconsistent with the model.  

3) \textit{The SZE-optical positional offset distribution.} 
We identify optical positional biases associated with 12\% of the sample due to the masking in the DES-SVA1 data. We remove these clusters and study the optical-SZE positional offset distribution for the rest of the matched sample. We model the underlying positional offset distribution as the sum of two
Gaussians, while accounting for the SPT-SZ positional uncertainty.  We show that
the resulting distribution is consistent with being described by a dominant ($63^{+15}_{-25}
\%$) centrally peaked distribution with ($\sigma_0 =
0.07^{+0.03}_{-0.02}$ \rvir) and a sub-dominant ($\sim 37\%$)
population characterized by larger separations ($\sigma_1 =
0.25^{+0.07}_{-0.06}$ \rvir). For the same population, the RM algorithm assumes that 79\% of the clusters will belong to a small-offset population, consistent with our observations. 

We also match the SPT-SZ cluster candidates with $4<\xi<4.5$ to the RM optical cluster catalogs from DES-SVA1 to extend the mass range of the SZE-selected clusters. Including the SPT-SZ candidates between $\xi=4$ and $\xi=4.5$ increases the sample of matched clusters by $\sim 40\%$ compared to the $\xi>4.5$ sample, highlighting the potential synergies of SPT and DES in producing lower-mass extensions of SZE-selected cluster samples.  We show that this larger sample produces results that are broadly consistent with the $\xi>4.5$ results, but only marginally tighter. This is due to the fact that mass constraints from lower signal-to-noise SPT clusters are somewhat weaker on a per cluster basis compared to the higher $\xi$ sample.
Future work benefiting from the larger region of overlap between the DES and SPT surveys will improve our derived constraints and help to better characterize the optical and SZE properties of cluster samples in terms of positional offsets, purity, and completeness.  Moreover, the multiwavelength datasets available through DES and SPT enable characterization of the galaxy populations of large SZE-selected cluster samples, calibration of the SZE-selected cluster masses using weak lensing constraints, and many other promising studies.

\section*{ACKNOWLEDGEMENTS}
We acknowledge the support by the DFG Cluster of Excellence "Origin
and Structure of the Universe", the Transregio program TR33 "The Dark
Universe" and the Ludwig-Maximilians University.  The South Pole
Telescope is supported by the National Science Foundation through
grant PLR-1248097. Partial support is also provided by the NSF Physics
Frontier Center grant PHY-1125897 to the Kavli Institute of
Cosmological Physics at the University of Chicago, the Kavli
Foundation and the Gordon and Betty Moore Foundation grant GBMF
947. A.A.S. acknowledges a Pell grant from the Smithsonian Institution.
T.~de~Haan is supported by a Miller Research Fellowship.
This work was partially completed at Fermilab, operated by Fermi
Research Alliance, LLC under Contract No. De-AC02-07CH11359 with the
United States Department of Energy.

\bibliographystyle{mn2e}
\bibliography{paper,spt}

\begin{thebibliography}{79}
\expandafter\ifx\csname natexlab\endcsname\relax\def\natexlab#1{#1}\fi

\bibitem[{{Abell}(1958)}]{Abell58}
{Abell} G.~O., 1958, \apjs, 3, 211

\bibitem[{{Aihara} {et~al}\mbox{.}(2011){Aihara}, {Allende Prieto}, {An},
  {Anderson}, {Aubourg}, {Balbinot}, {Beers}, {Berlind}, {Bickerton},
  {Bizyaev}, {Blanton}, {Bochanski}, {Bolton}, {Bovy}, {Brandt}, {Brinkmann},
  {Brown}, {Brownstein}, {Busca}, {Campbell}, {Carr}, {Chen}, {Chiappini},
  {Comparat}, {Connolly}, {Cortes}, {Croft}, {Cuesta}, {da Costa}, {Davenport},
  {Dawson}, {Dhital}, {Ealet}, {Ebelke}, {Edmondson}, {Eisenstein},
  {Escoffier}, {Esposito}, {Evans}, {Fan}, {Femen{\'{\i}}a Castell{\'a}},
  {Font-Ribera}, {Frinchaboy}, {Ge}, {Gillespie}, {Gilmore}, {Gonz{\'a}lez
  Hern{\'a}ndez}, {Gott}, {Gould}, {Grebel}, {Gunn}, {Hamilton}, {Harding},
  {Harris}, {Hawley}, {Hearty}, {Ho}, {Hogg}, {Holtzman}, {Honscheid}, {Inada},
  {Ivans}, {Jiang}, {Johnson}, {Jordan}, {Jordan}, {Kazin}, {Kirkby}, {Klaene},
  {Knapp}, {Kneib}, {Kochanek}, {Koesterke}, {Kollmeier}, {Kron}, {Lampeitl},
  {Lang}, {Le Goff}, {Lee}, {Lin}, {Long}, {Loomis}, {Lucatello}, {Lundgren},
  {Lupton}, {Ma}, {MacDonald}, {Mahadevan}, {Maia}, {Makler}, {Malanushenko},
  {Malanushenko}, {Mandelbaum}, {Maraston}, {Margala}, {Masters}, {McBride},
  {McGehee}, {McGreer}, {M{\'e}nard}, {Miralda-Escud{\'e}}, {Morrison},
  {Mullally}, {Muna}, {Munn}, {Murayama}, {Myers}, {Naugle}, {Neto}, {Nguyen},
  {Nichol}, {O'Connell}, {Ogando}, {Olmstead}, {Oravetz}, {Padmanabhan},
  {Palanque-Delabrouille}, {Pan}, {Pandey}, {P{\^a}ris}, {Percival},
  {Petitjean}, {Pfaffenberger}, {Pforr}, {Phleps}, {Pichon}, {Pieri}, {Prada},
  {Price-Whelan}, {Raddick}, {Ramos}, {Reyl{\'e}}, {Rich}, {Richards}, {Rix},
  {Robin}, {Rocha-Pinto}, {Rockosi}, {Roe}, {Rollinde}, {Ross}, {Ross},
  {Rossetto}, {S{\'a}nchez}, {Sayres}, {Schlegel}, {Schlesinger}, {Schmidt},
  {Schneider}, {Sheldon}, {Shu}, {Simmerer}, {Simmons}, {Sivarani}, {Snedden},
  {Sobeck}, {Steinmetz}, {Strauss}, {Szalay}, {Tanaka}, {Thakar}, {Thomas},
  {Tinker}, {Tofflemire}, {Tojeiro}, {Tremonti}, {Vandenberg}, {Vargas
  Maga{\~n}a}, {Verde}, {Vogt}, {Wake}, {Wang}, {Weaver}, {Weinberg}, {White},
  {White}, {Yanny}, {Yasuda}, {Yeche}, \& {Zehavi}}]{aihara11}
{Aihara} H. {et~al.}, 2011, \apjs, 193, 29

\bibitem[{{Allen} {et~al}\mbox{.}(2011){Allen}, {Evrard}, \& {Mantz}}]{allen11}
{Allen} S.~W., {Evrard} A.~E., {Mantz} A.~B., 2011, \araa, 49, 409

\bibitem[{{Andersson} {et~al}\mbox{.}(2011){Andersson}, {Benson}, {Ade},
  {Aird}, {Armstrong}, {Bautz}, {Bleem}, {Brodwin}, {Carlstrom}, {Chang},
  {Crawford}, {Crites}, {de Haan}, {Desai}, {Dobbs}, {Dudley}, {Foley},
  {Forman}, {Garmire}, {George}, {Gladders}, {Halverson}, {High}, {Holder},
  {Holzapfel}, {Hrubes}, {Jones}, {Joy}, {Keisler}, {Knox}, {Lee}, {Leitch},
  {Lueker}, {Marrone}, {McMahon}, {Mehl}, {Meyer}, {Mohr}, {Montroy}, {Murray},
  {Padin}, {Plagge}, {Pryke}, {Reichardt}, {Rest}, {Ruel}, {Ruhl}, {Schaffer},
  {Shaw}, {Shirokoff}, {Song}, {Spieler}, {Stalder}, {Staniszewski}, {Stark},
  {Stubbs}, {Vanderlinde}, {Vieira}, {Vikhlinin}, {Williamson}, {Yang}, {Zahn},
  \& {Zenteno}}]{andersson11}
{Andersson} K. {et~al.}, 2011, \apj, 738, 48

\bibitem[{{Andreon} \& {Congdon}(2014)}]{andreon14}
{Andreon} S., {Congdon} P., 2014, \aap, 568, A23

\bibitem[{{Annis} {et~al}\mbox{.}(2014){Annis}, {Soares-Santos}, {Strauss},
  {Becker}, {Dodelson}, {Fan}, {Gunn}, {Hao}, {Ivezi{\'c}}, {Jester}, {Jiang},
  {Johnston}, {Kubo}, {Lampeitl}, {Lin}, {Lupton}, {Miknaitis}, {Seo}, {Simet},
  \& {Yanny}}]{annis14}
{Annis} J. {et~al.}, 2014, \apj, 794, 120

\bibitem[{{Ascaso} {et~al}\mbox{.}(2014){Ascaso}, {Wittman}, \&
  {Dawson}}]{ascaso14}
{Ascaso} B., {Wittman} D., {Dawson} W., 2014, \mnras, 439, 1980

\bibitem[{{Banerji} {et~al}\mbox{.}(2015){Banerji}, {Jouvel}, {Lin}, {McMahon},
  {Lahav}, {Castander}, {Abdalla}, {Bertin}, {Bosman}, {Carnero}, {Kind}, {da
  Costa}, {Gerdes}, {Gschwend}, {Lima}, {Maia}, {Merson}, {Miller}, {Ogando},
  {Pellegrini}, {Reed}, {Saglia}, {S{\'a}nchez}, {Allam}, {Annis}, {Bernstein},
  {Bernstein}, {Bernstein}, {Capozzi}, {Childress}, {Cunha}, {Davis}, {DePoy},
  {Desai}, {Diehl}, {Doel}, {Findlay}, {Finley}, {Flaugher}, {Frieman},
  {Gaztanaga}, {Glazebrook}, {Gonz{\'a}lez-Fern{\'a}ndez}, {Gonzalez-Solares},
  {Honscheid}, {Irwin}, {Jarvis}, {Kim}, {Koposov}, {Kuehn}, {Kupcu-Yoldas},
  {Lagattuta}, {Lewis}, {Lidman}, {Makler}, {Marriner}, {Marshall}, {Miquel},
  {Mohr}, {Neilsen}, {Peoples}, {Sako}, {Sanchez}, {Scarpine}, {Schindler},
  {Schubnell}, {Sevilla}, {Sharp}, {Soares-Santos}, {Swanson}, {Tarle},
  {Thaler}, {Tucker}, {Uddin}, {Wechsler}, {Wester}, {Yuan}, \&
  {Zuntz}}]{banerji14}
{Banerji} M. {et~al.}, 2015, \mnras, 446, 2523

\bibitem[{{Bartlett} \& {Silk}(1994)}]{bartlett94}
{Bartlett} J.~G., {Silk} J., 1994, \apj, 423, 12

\bibitem[{{Benson} {et~al}\mbox{.}(2013){Benson}, {de Haan}, {Dudley},
  {Reichardt}, {Aird}, {Andersson}, {Armstrong}, {Ashby}, {Bautz}, {Bayliss},
  {Bazin}, {Bleem}, {Brodwin}, {Carlstrom}, {Chang}, {Cho}, {Clocchiatti},
  {Crawford}, {Crites}, {Desai}, {Dobbs}, {Foley}, {Forman}, {George},
  {Gladders}, {Gonzalez}, {Halverson}, {Harrington}, {High}, {Holder},
  {Holzapfel}, {Hoover}, {Hrubes}, {Jones}, {Joy}, {Keisler}, {Knox}, {Lee},
  {Leitch}, {Liu}, {Lueker}, {Luong-Van}, {Mantz}, {Marrone}, {McDonald},
  {McMahon}, {Mehl}, {Meyer}, {Mocanu}, {Mohr}, {Montroy}, {Murray}, {Natoli},
  {Padin}, {Plagge}, {Pryke}, {Rest}, {Ruel}, {Ruhl}, {Saliwanchik}, {Saro},
  {Sayre}, {Schaffer}, {Shaw}, {Shirokoff}, {Song}, {Spieler}, {Stalder},
  {Staniszewski}, {Stark}, {Story}, {Stubbs}, {Suhada}, {van Engelen},
  {Vanderlinde}, {Vieira}, {Vikhlinin}, {Williamson}, {Zahn}, \&
  {Zenteno}}]{benson11}
{Benson} B.~A. {et~al.}, 2013, \apj, 763, 147

\bibitem[{{Biesiadzinski} {et~al}\mbox{.}(2012){Biesiadzinski}, {McMahon},
  {Miller}, {Nord}, \& {Shaw}}]{biesiadzinski12}
{Biesiadzinski} T., {McMahon} J., {Miller} C.~J., {Nord} B., {Shaw} L., 2012,
  \apj, 757, 1

\bibitem[{{Bleem} {et~al}\mbox{.}(2015{\natexlab{a}}){Bleem}, {Stalder},
  {Brodwin}, {Busha}, {Gladders}, {High}, {Rest}, \& {Wechsler}}]{bleem14b}
{Bleem} L.~E., {Stalder} B., {Brodwin} M., {Busha} M.~T., {Gladders} M.~D.,
  {High} F.~W., {Rest} A., {Wechsler} R.~H., 2015{\natexlab{a}}, \apjs, 216, 20

\bibitem[{{Bleem} {et~al}\mbox{.}(2015{\natexlab{b}}){Bleem}, {Stalder}, {de
  Haan}, {Aird}, {Allen}, {Applegate}, {Ashby}, {Bautz}, {Bayliss}, {Benson},
  {Bocquet}, {Brodwin}, {Carlstrom}, {Chang}, {Chiu}, {Cho}, {Clocchiatti},
  {Crawford}, {Crites}, {Desai}, {Dietrich}, {Dobbs}, {Foley}, {Forman},
  {George}, {Gladders}, {Gonzalez}, {Halverson}, {Hennig}, {Hoekstra},
  {Holder}, {Holzapfel}, {Hrubes}, {Jones}, {Keisler}, {Knox}, {Lee}, {Leitch},
  {Liu}, {Lueker}, {Luong-Van}, {Mantz}, {Marrone}, {McDonald}, {McMahon},
  {Meyer}, {Mocanu}, {Mohr}, {Murray}, {Padin}, {Pryke}, {Reichardt}, {Rest},
  {Ruel}, {Ruhl}, {Saliwanchik}, {Saro}, {Sayre}, {Schaffer}, {Schrabback},
  {Shirokoff}, {Song}, {Spieler}, {Stanford}, {Staniszewski}, {Stark}, {Story},
  {Stubbs}, {Vanderlinde}, {Vieira}, {Vikhlinin}, {Williamson}, {Zahn}, \&
  {Zenteno}}]{bleem15}
{Bleem} L.~E. {et~al.}, 2015{\natexlab{b}}, \apjs, 216, 27

\bibitem[{{Bocquet} {et~al}\mbox{.}(2015){Bocquet}, {Saro}, {Mohr}, {Aird},
  {Ashby}, {Bautz}, {Bayliss}, {Bazin}, {Benson}, {Bleem}, {Brodwin},
  {Carlstrom}, {Chang}, {Chiu}, {Cho}, {Clocchiatti}, {Crawford}, {Crites},
  {Desai}, {de Haan}, {Dietrich}, {Dobbs}, {Foley}, {Forman}, {Gangkofner},
  {George}, {Gladders}, {Gonzalez}, {Halverson}, {Hennig}, {Hlavacek-Larrondo},
  {Holder}, {Holzapfel}, {Hrubes}, {Jones}, {Keisler}, {Knox}, {Lee}, {Leitch},
  {Liu}, {Lueker}, {Luong-Van}, {Marrone}, {McDonald}, {McMahon}, {Meyer},
  {Mocanu}, {Murray}, {Padin}, {Pryke}, {Reichardt}, {Rest}, {Ruel}, {Ruhl},
  {Saliwanchik}, {Sayre}, {Schaffer}, {Shirokoff}, {Spieler}, {Stalder},
  {Stanford}, {Staniszewski}, {Stark}, {Story}, {Stubbs}, {Vanderlinde},
  {Vieira}, {Vikhlinin}, {Williamson}, {Zahn}, \& {Zenteno}}]{bocquet15}
{Bocquet} S. {et~al.}, 2015, \apj, 799, 214

\bibitem[{{B{\"o}hringer} {et~al}\mbox{.}(2000){B{\"o}hringer}, {Voges},
  {Huchra}, {McLean}, {Giacconi}, {Rosati}, {Burg}, {Mader}, {Schuecker},
  {Simi{\c c}}, {Komossa}, {Reiprich}, {Retzlaff}, \&
  {Tr{\"u}mper}}]{boehringer00}
{B{\"o}hringer} H. {et~al.}, 2000, \apjs, 129, 435

\bibitem[{{Borgani} {et~al}\mbox{.}(2001){Borgani}, {Rosati}, {Tozzi},
  {Stanford}, {Eisenhardt}, {Lidman}, {Holden}, {Della Ceca}, {Norman}, \&
  {Squires}}]{borgani01}
{Borgani} S. {et~al.}, 2001, \apj, 561, 13

\bibitem[{{Cavaliere} \& {Fusco-Femiano}(1976)}]{cavaliere76}
{Cavaliere} A., {Fusco-Femiano} R., 1976, \aap, 49, 137

\bibitem[{{Crawford} {et~al}\mbox{.}(2010){Crawford}, {Switzer}, {Holzapfel},
  {Reichardt}, {Marrone}, \& {Vieira}}]{crawford10}
{Crawford} T.~M., {Switzer} E.~R., {Holzapfel} W.~L., {Reichardt} C.~L.,
  {Marrone} D.~P., {Vieira} J.~D., 2010, \apj, 718, 513

\bibitem[{{de Jong} {et~al}\mbox{.}(2013){de Jong}, {Verdoes Kleijn},
  {Kuijken}, \& {Valentijn}}]{kids}
{de Jong} J.~T.~A., {Verdoes Kleijn} G.~A., {Kuijken} K.~H., {Valentijn} E.~A.,
  2013, Experimental Astronomy, 35, 25

\bibitem[{{Desai} {et~al}\mbox{.}(2012){Desai}, {Armstrong}, {Mohr}, {Semler},
  {Liu}, {Bertin}, {Allam}, {Barkhouse}, {Bazin}, {Buckley-Geer}, {Cooper},
  {Hansen}, {High}, {Lin}, {Lin}, {Ngeow}, {Rest}, {Song}, {Tucker}, \&
  {Zenteno}}]{desai12}
{Desai} S. {et~al.}, 2012, \apj, 757, 83

\bibitem[{{Diehl T. et al.}(2012)}]{diehl12}
{Diehl T. et al.}, 2012, Physics Procedia, 37, 1332

\bibitem[{{Eisenhardt} {et~al}\mbox{.}(2008){Eisenhardt}, {Brodwin},
  {Gonzalez}, {Stanford}, {Stern}, {Barmby}, {Brown}, {Dawson}, {Dey}, {Doi},
  {Galametz}, {Jannuzi}, {Kochanek}, {Meyers}, {Morokuma}, \&
  {Moustakas}}]{eisenhardt08}
{Eisenhardt} P.~R.~M. {et~al.}, 2008, \apj, 684, 905

\bibitem[{{Eke} {et~al}\mbox{.}(1998){Eke}, {Cole}, {Frenk}, \& {Patrick
  Henry}}]{eke98}
{Eke} V.~R., {Cole} S., {Frenk} C.~S., {Patrick Henry} J., 1998, \mnras, 298,
  1145

\bibitem[{{Evrard} {et~al}\mbox{.}(2014){Evrard}, {Arnault}, {Huterer}, \&
  {Farahi}}]{evrard14}
{Evrard} A.~E., {Arnault} P., {Huterer} D., {Farahi} A., 2014, \mnras, 441,
  3562

\bibitem[{{Flaugher} {et~al}\mbox{.}(2015){Flaugher}, {Diehl}, {Honscheid},
  {Abbott}, {Alvarez}, {Angstadt}, {Annis}, {Antonik}, {Ballester}, {Beaufore},
  {Bernstein}, {Bernstein}, {Bigelow}, {Bonati}, {Boprie}, {Brooks},
  {Buckley-Geer}, {Campa}, {Cardiel-Sas}, {Castander}, {Castilla}, {Cease},
  {Cela-Ruiz}, {Chappa}, {Chi}, {Cooper}, {da Costa}, {Dede}, {Derylo},
  {DePoy}, {de Vicente}, {Doel}, {Drlica-Wagner}, {Eiting}, {Elliott}, {Emes},
  {Estrada}, {Fausti Neto}, {Finley}, {Flores}, {Frieman}, {Gerdes},
  {Gladders}, {Gregory}, {Gutierrez}, {Hao}, {Holland}, {Holm}, {Huffman},
  {Jackson}, {James}, {Jonas}, {Karcher}, {Karliner}, {Kent}, {Kessler},
  {Kozlovsky}, {Kron}, {Kubik}, {Kuehn}, {Kuhlmann}, {Kuk}, {Lahav}, {Lathrop},
  {Lee}, {Levi}, {Lewis}, {Li}, {Mandrichenko}, {Marshall}, {Martinez},
  {Merritt}, {Miquel}, {Munoz}, {Neilsen}, {Nichol}, {Nord}, {Ogando}, {Olsen},
  {Palio}, {Patton}, {Peoples}, {Plazas}, {Rauch}, {Reil}, {Rheault}, {Roe},
  {Rogers}, {Roodman}, {Sanchez}, {Scarpine}, {Schindler}, {Schmidt},
  {Schmitt}, {Schubnell}, {Schultz}, {Schurter}, {Scott}, {Serrano}, {Shaw},
  {Smith}, {Soares-Santos}, {Stefanik}, {Stuermer}, {Suchyta}, {Sypniewski},
  {Tarle}, {Thaler}, {Tighe}, {Tran}, {Tucker}, {Walker}, {Wang}, {Watson},
  {Weaverdyck}, {Wester}, {Woods}, \& {Yanny}}]{flaugher15}
{Flaugher} B. {et~al.}, 2015, ArXiv e-prints:1504.02900

\bibitem[{{Flaugher} {et~al}\mbox{.}(2012){Flaugher}, {Abbott}, {Angstadt},
  {Annis}, {Antonik}, {Bailey}, {Ballester}, {Bernstein}, {Bernstein},
  {Bonati}, {Bremer}, {Briones}, {Brooks}, {Buckley-Geer}, {Campa},
  {Cardiel-Sas}, {Castander}, {Castilla}, {Cease}, {Chappa}, {Chi}, {da Costa},
  {DePoy}, {Derylo}, {de Vincente}, {Diehl}, {Doel}, {Estrada}, {Eiting},
  {Elliott}, {Finley}, {Flores}, {Frieman}, {Gaztanaga}, {Gerdes}, {Gladders},
  {Guarino}, {Gutierrez}, {Grudzinski}, {Hanlon}, {Hao}, {Holland},
  {Honscheid}, {Huffman}, {Jackson}, {Jonas}, {Karliner}, {Kau}, {Kent},
  {Kozlovsky}, {Krempetz}, {Krider}, {Kubik}, {Kuehn}, {Kuhlmann}, {Kuk},
  {Lahav}, {Langellier}, {Lathrop}, {Lewis}, {Lin}, {Lorenzon}, {Martinez},
  {McKay}, {Merritt}, {Meyer}, {Miquel}, {Morgan}, {Moore}, {Moore}, {Neilsen},
  {Nord}, {Ogando}, {Olson}, {Patton}, {Peoples}, {Plazas}, {Qian}, {Roe},
  {Roodman}, {Rossetto}, {Sanchez}, {Soares-Santos}, {Scarpine}, {Schalk},
  {Schindler}, {Schmidt}, {Schmitt}, {Schubnell}, {Schultz}, {Selen},
  {Serrano}, {Shaw}, {Simaitis}, {Slaughter}, {Smith}, {Spinka}, {Stefanik},
  {Stuermer}, {Sypniewski}, {Talaga}, {Tarle}, {Thaler}, {Tucker}, {Walker},
  {Weaverdyck}, {Wester}, {Woods}, {Worswick}, \& {Zhao}}]{flaugher12}
{Flaugher} B.~L. {et~al.}, 2012, in Society of Photo-Optical Instrumentation
  Engineers (SPIE) Conference Series, Vol. 8446, Society of Photo-Optical
  Instrumentation Engineers (SPIE) Conference Series, p.~11

\bibitem[{{Gioia} {et~al}\mbox{.}(1990){Gioia}, {Maccacaro}, {Schild},
  {Wolter}, {Stocke}, {Morris}, \& {Henry}}]{gioia90}
{Gioia} I.~M., {Maccacaro} T., {Schild} R.~E., {Wolter} A., {Stocke} J.~T.,
  {Morris} S.~L., {Henry} J.~P., 1990, \apjs, 72, 567

\bibitem[{{Gladders} \& {Yee}(2000)}]{gladders00}
{Gladders} M.~D., {Yee} H. K.~C., 2000, \aj, 120, 2148

\bibitem[{{Haehnelt} \& {Tegmark}(1996)}]{haehnelt96}
{Haehnelt} M.~G., {Tegmark} M., 1996, \mnras, 279, 545+

\bibitem[{{Hao} {et~al}\mbox{.}(2010){Hao}, {McKay}, {Koester}, {Rykoff},
  {Rozo}, {Annis}, {Wechsler}, {Evrard}, {Siegel}, {Becker}, {Busha}, {Gerdes},
  {Johnston}, \& {Sheldon}}]{hao10}
{Hao} J. {et~al.}, 2010, \apjs, 191, 254

\bibitem[{{Hasselfield} {et~al}\mbox{.}(2013){Hasselfield}, {Hilton},
  {Marriage}, {Addison}, {Barrientos}, {Battaglia}, {Battistelli}, {Bond},
  {Crichton}, {Das}, {Devlin}, {Dicker}, {Dunkley}, {D{\"u}nner}, {Fowler},
  {Gralla}, {Hajian}, {Halpern}, {Hincks}, {Hlozek}, {Hughes}, {Infante},
  {Irwin}, {Kosowsky}, {Marsden}, {Menanteau}, {Moodley}, {Niemack}, {Nolta},
  {Page}, {Partridge}, {Reese}, {Schmitt}, {Sehgal}, {Sherwin}, {Sievers},
  {Sif{\'o}n}, {Spergel}, {Staggs}, {Swetz}, {Switzer}, {Thornton}, {Trac}, \&
  {Wollack}}]{hasselfield13}
{Hasselfield} M. {et~al.}, 2013, JCAP, 7, 8

\bibitem[{{High} {et~al}\mbox{.}(2012){High}, {Hoekstra}, {Leethochawalit}, {de
  Haan}, {Abramson}, {Aird}, {Armstrong}, {Ashby}, {Bautz}, {Bayliss}, {Bazin},
  {Benson}, {Bleem}, {Brodwin}, {Carlstrom}, {Chang}, {Cho}, {Clocchiatti},
  {Conroy}, {Crawford}, {Crites}, {Desai}, {Dobbs}, {Dudley}, {Foley},
  {Forman}, {George}, {Gladders}, {Gonzalez}, {Halverson}, {Harrington},
  {Holder}, {Holzapfel}, {Hoover}, {Hrubes}, {Jones}, {Joy}, {Keisler}, {Knox},
  {Lee}, {Leitch}, {Liu}, {Lueker}, {Luong-Van}, {Mantz}, {Marrone},
  {McDonald}, {McMahon}, {Mehl}, {Meyer}, {Mocanu}, {Mohr}, {Montroy},
  {Murray}, {Natoli}, {Nurgaliev}, {Padin}, {Plagge}, {Pryke}, {Reichardt},
  {Rest}, {Ruel}, {Ruhl}, {Saliwanchik}, {Saro}, {Sayre}, {Schaffer}, {Shaw},
  {Schrabback}, {Shirokoff}, {Song}, {Spieler}, {Stalder}, {Staniszewski},
  {Stark}, {Story}, {Stubbs}, {{\v S}uhada}, {Tokarz}, {van Engelen},
  {Vanderlinde}, {Vieira}, {Vikhlinin}, {Williamson}, {Zahn}, \&
  {Zenteno}}]{high12}
{High} F.~W. {et~al.}, 2012, \apj, 758, 68

\bibitem[{{Koester} {et~al}\mbox{.}(2007){Koester}, {McKay}, {Annis},
  {Wechsler}, {Evrard}, {Bleem}, {Becker}, {Johnston}, {Sheldon}, {Nichol},
  {Miller}, {Scranton}, {Bahcall}, {Barentine}, {Brewington}, {Brinkmann},
  {Harvanek}, {Kleinman}, {Krzesinski}, {Long}, {Nitta}, {Schneider},
  {Sneddin}, {Voges}, \& {York}}]{koester07a}
{Koester} B.~P. {et~al.}, 2007, \apj, 660, 239

\bibitem[{{Lauer} {et~al}\mbox{.}(2014){Lauer}, {Postman}, {Strauss}, {Graves},
  \& {Chisari}}]{lauer14}
{Lauer} T.~R., {Postman} M., {Strauss} M.~A., {Graves} G.~J., {Chisari} N.~E.,
  2014, \apj, 797, 82

\bibitem[{{Laureijs} {et~al}\mbox{.}(2011){Laureijs}, {Amiaux}, {Arduini},
  {Augu{\`e}res}, {Brinchmann}, {Cole}, {Cropper}, {Dabin}, {Duvet}, {Ealet},
  \& et~al.}]{euclid}
{Laureijs} R. {et~al.}, 2011, ArXiv e-prints:1110.3193

\bibitem[{{Lin} {et~al}\mbox{.}(2004){Lin}, {Mohr}, \& {Stanford}}]{lin04a}
{Lin} Y., {Mohr} J.~J., {Stanford} S.~A., 2004, \apj, 610, 745

\bibitem[{{Liu} {et~al}\mbox{.}(2015){Liu}, {Mohr}, {Saro}, {Aird}, {Ashby},
  {Bautz}, {Bayliss}, {Benson}, {Bleem}, {Bocquet}, {Brodwin}, {Carlstrom},
  {Chang}, {Chiu}, {Cho}, {Clocchiatti}, {Crawford}, {Crites}, {de Haan},
  {Desai}, {Dietrich}, {Dobbs}, {Foley}, {Gangkofner}, {George}, {Gladders},
  {Gonzalez}, {Halverson}, {Hennig}, {Hlavacek-Larrondo}, {Holder},
  {Holzapfel}, {Hrubes}, {Jones}, {Keisler}, {Lee}, {Leitch}, {Lueker},
  {Luong-Van}, {McDonald}, {McMahon}, {Meyer}, {Mocanu}, {Murray}, {Padin},
  {Pryke}, {Reichardt}, {Rest}, {Ruel}, {Ruhl}, {Saliwanchik}, {Sayre},
  {Schaffer}, {Shirokoff}, {Spieler}, {Stalder}, {Staniszewski}, {Stark},
  {Story}, {{\v S}uhada}, {Vanderlinde}, {Vieira}, {Vikhlinin}, {Williamson},
  {Zahn}, \& {Zenteno}}]{liu14}
{Liu} J. {et~al.}, 2015, \mnras, 448, 2085

\bibitem[{{LSST Dark Energy Science Collaboration}(2012)}]{lsst}
{LSST Dark Energy Science Collaboration}, 2012, ArXiv e-prints:1211.0310

\bibitem[{{Mantz} {et~al}\mbox{.}(2010){Mantz}, {Allen}, {Ebeling}, {Rapetti},
  \& {Drlica-Wagner}}]{mantz10}
{Mantz} A., {Allen} S.~W., {Ebeling} H., {Rapetti} D., {Drlica-Wagner} A.,
  2010, \mnras, 406, 1773

\bibitem[{{Mei} {et~al}\mbox{.}(2009){Mei}, {Holden}, {Blakeslee}, {Ford},
  {Franx}, {Homeier}, {Illingworth}, {Jee}, {Overzier}, {Postman}, {Rosati},
  {Van der Wel}, \& {Bartlett}}]{mei09}
{Mei} S. {et~al.}, 2009, \apj, 690, 42

\bibitem[{{Melchior} {et~al}\mbox{.}(2015){Melchior}, {Suchyta}, {Huff},
  {Hirsch}, {Kacprzak}, {Rykoff}, {Gruen}, {Armstrong}, {Bacon}, {Bechtol},
  {Bernstein}, {Bridle}, {Clampitt}, {Honscheid}, {Jain}, {Jouvel}, {Krause},
  {Lin}, {MacCrann}, {Patton}, {Plazas}, {Rowe}, {Vikram}, {Wilcox}, {Young},
  {Zuntz}, {Abbott}, {Abdalla}, {Allam}, {Banerji}, {Bernstein}, {Bernstein},
  {Bertin}, {Buckley-Geer}, {Burke}, {Castander}, {da Costa}, {Cunha}, {Depoy},
  {Desai}, {Diehl}, {Doel}, {Estrada}, {Evrard}, {Neto}, {Fernandez}, {Finley},
  {Flaugher}, {Frieman}, {Gaztanaga}, {Gerdes}, {Gruendl}, {Gutierrez},
  {Jarvis}, {Karliner}, {Kent}, {Kuehn}, {Kuropatkin}, {Lahav}, {Maia},
  {Makler}, {Marriner}, {Marshall}, {Merritt}, {Miller}, {Miquel}, {Mohr},
  {Neilsen}, {Nichol}, {Nord}, {Reil}, {Roe}, {Roodman}, {Sako}, {Sanchez},
  {Santiago}, {Schindler}, {Schubnell}, {Sevilla-Noarbe}, {Sheldon}, {Smith},
  {Soares-Santos}, {Swanson}, {Sypniewski}, {Tarle}, {Thaler}, {Thomas},
  {Tucker}, {Walker}, {Wechsler}, {Weller}, \& {Wester}}]{melchior15}
{Melchior} P. {et~al.}, 2015, \mnras, 449, 2219

\bibitem[{{Melin} {et~al}\mbox{.}(2006){Melin}, {Bartlett}, \&
  {Delabrouille}}]{melin06}
{Melin} J.-B., {Bartlett} J.~G., {Delabrouille} J., 2006, \aap, 459, 341

\bibitem[{{Menanteau} {et~al}\mbox{.}(2010){Menanteau}, {Hughes}, {Barrientos},
  {Deshpande}, {Hilton}, {Infante}, {Jimenez}, {Kosowsky}, {Moodley},
  {Spergel}, \& {Verde}}]{menanteau10}
{Menanteau} F. {et~al.}, 2010, \apjs, 191, 340

\bibitem[{{Mohr} {et~al}\mbox{.}(2008){Mohr}, {Adams}, {Barkhouse}, {Beldica},
  {Bertin}, {Cai}, {da Costa}, {Darnell}, {Daues}, {Jarvis}, {Gower}, {Lin},
  {Martelli}, {Neilsen}, {Ngeow}, {Ogando}, {Parga}, {Sheldon}, {Tucker},
  {Kuropatkin}, \& {Stoughton}}]{mohr08}
{Mohr} J.~J. {et~al.}, 2008, in Society of Photo-Optical Instrumentation
  Engineers (SPIE) Conference Series, Vol. 7016, Society of Photo-Optical
  Instrumentation Engineers (SPIE) Conference Series

\bibitem[{{Mohr} {et~al}\mbox{.}(2012){Mohr}, {Armstrong}, {Bertin}, {Daues},
  {Desai}, {Gower}, {Gruendl}, {Hanlon}, {Kuropatkin}, {Lin}, {Marriner},
  {Petravic}, {Sevilla}, {Swanson}, {Tomashek}, {Tucker}, \& {Yanny}}]{mohr12}
{Mohr} J.~J. {et~al.}, 2012, in Society of Photo-Optical Instrumentation
  Engineers (SPIE) Conference Series, Vol. 8451, Society of Photo-Optical
  Instrumentation Engineers (SPIE) Conference Series, p.~0

\bibitem[{{Moscovich-Eiger} {et~al}\mbox{.}(2013){Moscovich-Eiger}, {Nadler},
  \& {Spiegelman}}]{2013arXiv1311.3190M}
{Moscovich-Eiger} A., {Nadler} B., {Spiegelman} C., 2013, ArXiv e-prints

\bibitem[{{Muzzin} {et~al}\mbox{.}(2012){Muzzin}, {Wilson}, {Yee}, {Gilbank},
  {Hoekstra}, {Demarco}, {Balogh}, {van Dokkum}, {Franx}, {Ellingson}, {Hicks},
  {Nantais}, {Noble}, {Lacy}, {Lidman}, {Rettura}, {Surace}, \&
  {Webb}}]{muzzin12}
{Muzzin} A. {et~al.}, 2012, \apj, 746, 188

\bibitem[{{Ngeow} {et~al}\mbox{.}(2006){Ngeow}, {Mohr}, {Alam}, {Barkhouse},
  {Beldica}, {Cai}, {Daues}, {Plante}, {Annis}, {Lin}, {Tucker}, \&
  {Smith}}]{ngeow06}
{Ngeow} C. {et~al.}, 2006, in Society of Photo-Optical Instrumentation
  Engineers (SPIE) Conference Series, Vol. 6270, Society of Photo-Optical
  Instrumentation Engineers (SPIE) Conference Series

\bibitem[{{Pacaud} {et~al}\mbox{.}(2007){Pacaud}, {Pierre}, {Adami}, {Altieri},
  {Andreon}, {Chiappetti}, {Detal}, {Duc}, {Galaz}, {Gueguen}, {Le F{\`e}vre},
  {Hertling}, {Libbrecht}, {Melin}, {Ponman}, {Quintana}, {Refregier},
  {Sprimont}, {Surdej}, {Valtchanov}, {Willis}, {Alloin}, {Birkinshaw},
  {Bremer}, {Garcet}, {Jean}, {Jones}, {Le F{\`e}vre}, {Maccagni}, {Mazure},
  {Proust}, {R{\"o}ttgering}, \& {Trinchieri}}]{pacaud07}
{Pacaud} F. {et~al.}, 2007, \mnras, 382, 1289

\bibitem[{{Planck Collaboration} {et~al}\mbox{.}(2015){Planck Collaboration},
  {Ade}, {Aghanim}, {Arnaud}, {Ashdown}, {Aumont}, {Baccigalupi}, {Banday},
  {Barreiro}, {Barrena}, \& et~al.}]{planck15clusters}
{Planck Collaboration} {et~al.}, 2015, ArXiv e-prints

\bibitem[{{Reichardt} {et~al}\mbox{.}(2013){Reichardt}, {Stalder}, {Bleem},
  {Montroy}, {Aird}, {Andersson}, {Armstrong}, {Ashby}, {Bautz}, {Bayliss},
  {Bazin}, {Benson}, {Brodwin}, {Carlstrom}, {Chang}, {Cho}, {Clocchiatti},
  {Crawford}, {Crites}, {de Haan}, {Desai}, {Dobbs}, {Dudley}, {Foley},
  {Forman}, {George}, {Gladders}, {Gonzalez}, {Halverson}, {Harrington},
  {High}, {Holder}, {Holzapfel}, {Hoover}, {Hrubes}, {Jones}, {Joy}, {Keisler},
  {Knox}, {Lee}, {Leitch}, {Liu}, {Lueker}, {Luong-Van}, {Mantz}, {Marrone},
  {McDonald}, {McMahon}, {Mehl}, {Meyer}, {Mocanu}, {Mohr}, {Murray}, {Natoli},
  {Padin}, {Plagge}, {Pryke}, {Rest}, {Ruel}, {Ruhl}, {Saliwanchik}, {Saro},
  {Sayre}, {Schaffer}, {Shaw}, {Shirokoff}, {Song}, {Spieler}, {Staniszewski},
  {Stark}, {Story}, {Stubbs}, {{\v S}uhada}, {van Engelen}, {Vanderlinde},
  {Vieira}, {Vikhlinin}, {Williamson}, {Zahn}, \& {Zenteno}}]{reichardt13}
{Reichardt} C.~L. {et~al.}, 2013, \apj, 763, 127

\bibitem[{{Rozo} {et~al}\mbox{.}(2014{\natexlab{a}}){Rozo}, {Bartlett},
  {Evrard}, \& {Rykoff}}]{rozo14c}
{Rozo} E., {Bartlett} J.~G., {Evrard} A.~E., {Rykoff} E.~S.,
  2014{\natexlab{a}}, \mnras, 438, 78

\bibitem[{{Rozo} {et~al}\mbox{.}(2014{\natexlab{b}}){Rozo}, {Evrard}, {Rykoff},
  \& {Bartlett}}]{rozo14d}
{Rozo} E., {Evrard} A.~E., {Rykoff} E.~S., {Bartlett} J.~G.,
  2014{\natexlab{b}}, \mnras, 438, 62

\bibitem[{{Rozo} \& {Rykoff}(2014)}]{rozo14b}
{Rozo} E., {Rykoff} E.~S., 2014, \apj, 783, 80

\bibitem[{{Rozo} {et~al}\mbox{.}(2014{\natexlab{c}}){Rozo}, {Rykoff},
  {Bartlett}, \& {Melin}}]{rozo14a}
{Rozo} E., {Rykoff} E.~S., {Bartlett} J.~G., {Melin} J.~B., 2014{\natexlab{c}},
  ArXiv e-prints

\bibitem[{{Rozo} {et~al}\mbox{.}(2010){Rozo}, {Wechsler}, {Rykoff}, {Annis},
  {Becker}, {Evrard}, {Frieman}, {Hansen}, {Hao}, {Johnston}, {Koester},
  {McKay}, {Sheldon}, \& {Weinberg}}]{rozo10}
{Rozo} E. {et~al.}, 2010, \apj, 708, 645

\bibitem[{{Rozo et al.}(2015)}]{rozo15}
{Rozo et al.}, 2015, In Preparation

\bibitem[{{Rykoff} {et~al}\mbox{.}(2012){Rykoff}, {Koester}, {Rozo}, {Annis},
  {Evrard}, {Hansen}, {Hao}, {Johnston}, {McKay}, \& {Wechsler}}]{rykoff12}
{Rykoff} E.~S. {et~al.}, 2012, \apj, 746, 178

\bibitem[{{Rykoff} {et~al}\mbox{.}(2014){Rykoff}, {Rozo}, {Busha}, {Cunha},
  {Finoguenov}, {Evrard}, {Hao}, {Koester}, {Leauthaud}, {Nord}, {Pierre},
  {Reddick}, {Sadibekova}, {Sheldon}, \& {Wechsler}}]{rykoff14}
{Rykoff} E.~S. {et~al.}, 2014, \apj, 785, 104

\bibitem[{{Rykoff et al.}(2015)}]{rykoff15}
{Rykoff et al.}, 2015, In Preparation

\bibitem[{{S{\'a}nchez} {et~al}\mbox{.}(2014){S{\'a}nchez}, {Carrasco Kind},
  {Lin}, {Miquel}, {Abdalla}, {Amara}, {Banerji}, {Bonnett}, {Brunner},
  {Capozzi}, {Carnero}, {Castander}, {da Costa}, {Cunha}, {Fausti}, {Gerdes},
  {Greisel}, {Gschwend}, {Hartley}, {Jouvel}, {Lahav}, {Lima}, {Maia},
  {Mart{\'{\i}}}, {Ogando}, {Ostrovski}, {Pellegrini}, {Rau}, {Sadeh}, {Seitz},
  {Sevilla-Noarbe}, {Sypniewski}, {de Vicente}, {Abbot}, {Allam}, {Atlee},
  {Bernstein}, {Bernstein}, {Buckley-Geer}, {Burke}, {Childress}, {Davis},
  {DePoy}, {Dey}, {Desai}, {Diehl}, {Doel}, {Estrada}, {Evrard},
  {Fern{\'a}ndez}, {Finley}, {Flaugher}, {Frieman}, {Gaztanaga}, {Glazebrook},
  {Honscheid}, {Kim}, {Kuehn}, {Kuropatkin}, {Lidman}, {Makler}, {Marshall},
  {Nichol}, {Roodman}, {S{\'a}nchez}, {Santiago}, {Sako}, {Scalzo}, {Smith},
  {Swanson}, {Tarle}, {Thomas}, {Tucker}, {Uddin}, {Vald{\'e}s}, {Walker},
  {Yuan}, \& {Zuntz}}]{sanchez14}
{S{\'a}nchez} C. {et~al.}, 2014, \mnras, 445, 1482

\bibitem[{{Sehgal} {et~al}\mbox{.}(2013){Sehgal}, {Addison}, {Battaglia},
  {Battistelli}, {Bond}, {Das}, {Devlin}, {Dunkley}, {D{\"u}nner}, {Gralla},
  {Hajian}, {Halpern}, {Hasselfield}, {Hilton}, {Hincks}, {Hlozek}, {Hughes},
  {Kosowsky}, {Lin}, {Louis}, {Marriage}, {Marsden}, {Menanteau}, {Moodley},
  {Niemack}, {Page}, {Partridge}, {Reese}, {Sherwin}, {Sievers}, {Sif{\'o}n},
  {Spergel}, {Staggs}, {Swetz}, {Switzer}, \& {Wollack}}]{sehgal13}
{Sehgal} N. {et~al.}, 2013, \apj, 767, 38

\bibitem[{{Soares-Santos} {et~al}\mbox{.}(2011){Soares-Santos}, {de Carvalho},
  {Annis}, {Gal}, {La Barbera}, {Lopes}, {Wechsler}, {Busha}, \&
  {Gerke}}]{soares11}
{Soares-Santos} M. {et~al.}, 2011, \apj, 727, 45

\bibitem[{{Song} {et~al}\mbox{.}(2012){Song}, {Zenteno}, {Stalder}, {Desai},
  {Bleem}, {Aird}, {Armstrong}, {Ashby}, {Bayliss}, {Bazin}, {Benson},
  {Bertin}, {Brodwin}, {Carlstrom}, {Chang}, {Cho}, {Clocchiatti}, {Crawford},
  {Crites}, {de Haan}, {Dobbs}, {Dudley}, {Foley}, {George}, {Gettings},
  {Gladders}, {Gonzalez}, {Halverson}, {Harrington}, {High}, {Holder},
  {Holzapfel}, {Hoover}, {Hrubes}, {Joy}, {Keisler}, {Knox}, {Lee}, {Leitch},
  {Liu}, {Lueker}, {Luong-Van}, {Marrone}, {McDonald}, {McMahon}, {Mehl},
  {Meyer}, {Mocanu}, {Mohr}, {Montroy}, {Natoli}, {Nurgaliev}, {Padin},
  {Plagge}, {Pryke}, {Reichardt}, {Rest}, {Ruel}, {Ruhl}, {Saliwanchik},
  {Saro}, {Sayre}, {Schaffer}, {Shaw}, {Shirokoff}, {{\v S}uhada}, {Spieler},
  {Stanford}, {Staniszewski}, {Stark}, {Story}, {Stubbs}, {van Engelen},
  {Vanderlinde}, {Vieira}, {Williamson}, \& {Zahn}}]{song12}
{Song} J. {et~al.}, 2012, \apj, 761, 22

\bibitem[{{Staniszewski} {et~al}\mbox{.}(2009){Staniszewski}, {Ade}, {Aird},
  {Benson}, {Bleem}, {Carlstrom}, {Chang}, {Cho}, {Crawford}, {Crites}, {de
  Haan}, {Dobbs}, {Halverson}, {Holder}, {Holzapfel}, {Hrubes}, {Joy},
  {Keisler}, {Lanting}, {Lee}, {Leitch}, {Loehr}, {Lueker}, {McMahon}, {Mehl},
  {Meyer}, {Mohr}, {Montroy}, {Ngeow}, {Padin}, {Plagge}, {Pryke}, {Reichardt},
  {Ruhl}, {Schaffer}, {Shaw}, {Shirokoff}, {Spieler}, {Stalder}, {Stark},
  {Vanderlinde}, {Vieira}, {Zahn}, \& {Zenteno}}]{staniszewski09}
{Staniszewski} Z. {et~al.}, 2009, \apj, 701, 32

\bibitem[{{Story} {et~al}\mbox{.}(2011){Story}, {Aird}, {Andersson},
  {Armstrong}, {Bazin}, {Benson}, {Bleem}, {Bonamente}, {Brodwin}, {Carlstrom},
  {Chang}, {Clocchiatti}, {Crawford}, {Crites}, {de Haan}, {Desai}, {Dobbs},
  {Dudley}, {Foley}, {George}, {Gladders}, {Gonzalez}, {Halverson}, {High},
  {Holder}, {Holzapfel}, {Hoover}, {Hrubes}, {Joy}, {Keisler}, {Knox}, {Lee},
  {Leitch}, {Lueker}, {Luong-Van}, {Marrone}, {McMahon}, {Mehl}, {Meyer},
  {Mohr}, {Montroy}, {Padin}, {Plagge}, {Pryke}, {Reichardt}, {Rest}, {Ruel},
  {Ruhl}, {Saliwanchik}, {Saro}, {Schaffer}, {Shaw}, {Shirokoff}, {Song},
  {Spieler}, {Stalder}, {Staniszewski}, {Stark}, {Stubbs}, {Vanderlinde},
  {Vieira}, {Williamson}, \& {Zenteno}}]{story11}
{Story} K. {et~al.}, 2011, \apjl, 735, L36+

\bibitem[{{Sunyaev} \& {Zel'dovich}(1972)}]{Sunyaev72}
{Sunyaev} R.~A., {Zel'dovich} Y.~B., 1972, Comments on Astrophysics and Space
  Physics, 4, 173

\bibitem[{{The Dark Energy Survey Collaboration}(2005)}]{des}
{The Dark Energy Survey Collaboration}, 2005, ArXiv Astrophysics e-prints

\bibitem[{{Tinker} {et~al}\mbox{.}(2008){Tinker}, {Kravtsov}, {Klypin},
  {Abazajian}, {Warren}, {Yepes}, {Gottl{\"o}ber}, \& {Holz}}]{tinker08}
{Tinker} J., {Kravtsov} A.~V., {Klypin} A., {Abazajian} K., {Warren} M.,
  {Yepes} G., {Gottl{\"o}ber} S., {Holz} D.~E., 2008, \apj, 688, 709

\bibitem[{{{\v S}uhada} {et~al}\mbox{.}(2012){{\v S}uhada}, {Song},
  {B{\"o}hringer}, {Mohr}, {Chon}, {Finoguenov}, {Fassbender}, {Desai},
  {Armstrong}, {Zenteno}, {Barkhouse}, {Bertin}, {Buckley-Geer}, {Hansen},
  {High}, {Lin}, {M{\"u}hlegger}, {Ngeow}, {Pierini}, {Pratt}, {Verdugo}, \&
  {Tucker}}]{suhada12}
{{\v S}uhada} R. {et~al.}, 2012, \aap, 537, A39

\bibitem[{{Vanderlinde} {et~al}\mbox{.}(2010){Vanderlinde}, {Crawford}, {de
  Haan}, {Dudley}, {Shaw}, {Ade}, {Aird}, {Benson}, {Bleem}, {Brodwin},
  {Carlstrom}, {Chang}, {Crites}, {Desai}, {Dobbs}, {Foley}, {George},
  {Gladders}, {Hall}, {Halverson}, {High}, {Holder}, {Holzapfel}, {Hrubes},
  {Joy}, {Keisler}, {Knox}, {Lee}, {Leitch}, {Loehr}, {Lueker}, {Marrone},
  {McMahon}, {Mehl}, {Meyer}, {Mohr}, {Montroy}, {Ngeow}, {Padin}, {Plagge},
  {Pryke}, {Reichardt}, {Rest}, {Ruel}, {Ruhl}, {Schaffer}, {Shirokoff},
  {Song}, {Spieler}, {Stalder}, {Staniszewski}, {Stark}, {Stubbs}, {van
  Engelen}, {Vieira}, {Williamson}, {Yang}, {Zahn}, \&
  {Zenteno}}]{vanderlinde10}
{Vanderlinde} K. {et~al.}, 2010, \apj, 722, 1180

\bibitem[{Viana \& Liddle(1999)}]{viana99}
Viana P., Liddle A., 1999, \mnras, 303, 535

\bibitem[{{Vikhlinin} {et~al}\mbox{.}(2009){Vikhlinin}, {Kravtsov}, {Burenin},
  {Ebeling}, {Forman}, {Hornstrup}, {Jones}, {Murray}, {Nagai}, {Quintana}, \&
  {Voevodkin}}]{vikhlinin09}
{Vikhlinin} A. {et~al.}, 2009, \apj, 692, 1060

\bibitem[{Vikhlinin {et~al}\mbox{.}(1998)Vikhlinin, McNamara, Forman, Jones,
  Quintana, \& Hornstrup}]{vikhlinin98}
Vikhlinin A., McNamara B., Forman W., Jones C., Quintana H., Hornstrup A.,
  1998, \apj, 502, 558

\bibitem[{{Wen} {et~al}\mbox{.}(2012){Wen}, {Han}, \& {Liu}}]{wen12}
{Wen} Z.~L., {Han} J.~L., {Liu} F.~S., 2012, \apjs, 199, 34

\bibitem[{{White} {et~al}\mbox{.}(1993){White}, {Efstathiou}, \&
  {Frenk}}]{white93b}
{White} S.~D.~M., {Efstathiou} G., {Frenk} C.~S., 1993, \mnras, 262, 1023

\bibitem[{{Wiesner} {et~al}\mbox{.}(2015){Wiesner}, {Lin}, \&
  {Soares-Santos}}]{2015arXiv150106893W}
{Wiesner} M.~P., {Lin} H., {Soares-Santos} M., 2015, ArXiv e-prints

\bibitem[{{Zenteno} {et~al}\mbox{.}(2011){Zenteno}, {Song}, {Desai},
  {Armstrong}, {Mohr}, {Ngeow}, {Barkhouse}, {Allam}, {Andersson}, {Bazin},
  {Benson}, {Bertin}, {Brodwin}, {Buckley-Geer}, {Hansen}, {High}, {Lin},
  {Lin}, {Liu}, {Rest}, {Smith}, {Stalder}, {Stark}, {Tucker}, \&
  {Yang}}]{zenteno10}
{Zenteno} A. {et~al.}, 2011, \apj, 734, 3

\bibitem[{Zwicky {et~al}\mbox{.}(1968)Zwicky, Herzog, \& Wild}]{ZW68.1}
Zwicky F., Herzog E., Wild P., 1968, Catalogue of galaxies and of clusters of
  galaxies. California Institute of Technology, Pasadena

\end{thebibliography}
\textit{$^{1}$Department of Physics, Ludwig-Maximilians-Universitaet, Scheinerstr. 1, 81679 Muenchen, Germany\\
$^{2}$Excellence Cluster Universe, Boltzmannstr.\ 2, 85748 Garching, Germany\\
$^{3}$Department of Physics, University of Arizona, 1118 E 4th St, Tucson, AZ 85721\\
$^{4}$Fermi National Accelerator Laboratory, P. O. Box 500, Batavia, IL 60510, USA\\
$^{5}$Kavli Institute for Cosmological Physics, University of Chicago, Chicago, IL 60637, USA\\
$^{6}$Department of Astronomy and Astrophysics,University of Chicago, 5640 South Ellis Avenue, Chicago, IL 60637\\
$^{7}$Max Planck Institute for Extraterrestrial Physics, Giessenbachstrasse, 85748 Garching, Germany\\
$^{8}$Kavli Institute for Particle Astrophysics \& Cosmology, P. O. Box 2450, Stanford University, Stanford, CA 94305, USA\\
$^{9}$SLAC National Accelerator Laboratory, Menlo Park, CA 94025, USA\\
$^{10}$Argonne National Laboratory, 9700 South Cass Avenue, Lemont, IL 60439, USA\\
$^{11}$Center for Cosmology and Astro-Particle Physics, The Ohio State University, Columbus, OH 43210, USA\\
$^{12}$Department of Physics, The Ohio State University, Columbus, OH 43210, USA\\
$^{13}$Laborat\'orio Interinstitucional de e-Astronomia - LIneA, Rua Gal. Jos\'e Cristino 77, Rio de Janeiro, RJ - 20921-400, Brazil\\
$^{14}$CERN, CH-1211 Geneva 23, Switzerland\\
$^{15}$Universit\"ats-Sternwarte, Fakult\"at f\"ur Physik, Ludwig-Maximilians Universit\"at M\"unchen, Scheinerstr. 1, 81679 M\"unchen, Germany\\
$^{16}$Cerro Tololo Inter-American Observatory, National Optical Astronomy Observatory, Casilla 603, La Serena, Chile\\
$^{17}$Department of Physics \& Astronomy, University College London, Gower Street, London, WC1E 6BT, UK\\
$^{18}$Department of Physics and Astronomy, University of Pennsylvania, Philadelphia, PA 19104, USA\\
$^{19}$Institute of Astronomy, University of Cambridge, Madingley Road, Cambridge CB3 0HA, UK\\
$^{20}$Institut de Ci\`encies de l'Espai, IEEC-CSIC, Campus UAB, Facultat de Ci\`encies, Torre C5 par-2, 08193 Bellaterra, Barcelona, Spain\\
$^{21}$Department of Physics, Harvard University, 17 Oxford Street, Cambridge, MA 02138\\
$^{22}$Harvard-Smithsonian Center for Astrophysics, 60 Garden Street, Cambridge, MA 02138\\
$^{23}$Institut d'Astrophysique de Paris, Univ. Pierre et Marie Curie \& CNRS UMR7095, F-75014 Paris, France\\
$^{24}$Department of Physics and Astronomy, University of Missouri, 5110 Rockhill Road, Kansas City, MO 64110\\
$^{25}$ Department of Physics and Enrico Fermi Institute, University of Chicago, 5640 South Ellis Avenue, Chicago, IL 60637\\
$^{26}$Institute of Cosmology \& Gravitation, University of Portsmouth, Portsmouth, PO1 3FX, UK\\
$^{27}$Observat\'orio Nacional, Rua Gal. Jos\'e Cristino 77, Rio de Janeiro, RJ - 20921-400, Brazil\\
$^{28}$Department of Astronomy, University of Illinois,1002 W. Green Street, Urbana, IL 61801, USA\\
$^{29}$National Center for Supercomputing Applications, 1205 West Clark St., Urbana, IL 61801, USA\\
$^{30}$George P. and Cynthia Woods Mitchell Institute for Fundamental Physics and Astronomy, and Department of Physics and Astronomy, Texas A\&M University, College Station, TX 77843,  USA\\
$^{31}$Department of Physics, McGill University, 3600 Rue University, Montreal, Quebec H3A 2T8, Canada\\
$^{32}$Department of Physics, University of California, Berkeley, CA 94720\\
$^{33}$Jet Propulsion Laboratory, California Institute of Technology, 4800 Oak Grove Dr., Pasadena, CA 91109, USA\\
$^{34}$Department of Astronomy, University of Michigan, Ann Arbor, MI 48109, USA\\
$^{35}$Department of Physics, University of Michigan, Ann Arbor, MI 48109, USA\\
$^{36}$Institut de F\'{\i}sica d'Altes Energies, Universitat Aut\`onoma de Barcelona, E-08193 Bellaterra, Barcelona, Spain\\
$^{37}$Australian Astronomical Observatory, North Ryde, NSW 2113, Australia\\
$^{38}$Department of Astronomy, The Ohio State University, Columbus, OH 43210, USA\\
$^{39}$Kavli Institute for Astrophysics and Space Research, Massachusetts Institute of Technology, 77 Massachusetts Avenue, Cambridge, MA 02139\\
$^{40}$Brookhaven National Laboratory, Bldg 510, Upton, NY 11973, USA\\
$^{41}$School of Physics, University of Melbourne, Parkville, VIC 3010, Australia\\
$^{42}$Department of Physics and Astronomy, Pevensey Building, University of Sussex, Brighton, BN1 9QH, UK\\
$^{43}$Centro de Investigaciones Energ\'eticas, Medioambientales y Tecnol\'ogicas (CIEMAT), Madrid, Spain\\
$^{44}$Institute for Astronomy, University of Hawaii at Manoa, Honolulu, HI 96822, USA\\
$^{45}$Department of Physics, University of Illinois, 1110 W. Green St., Urbana, IL 61801, USA\\
$^{46}$Department of Physics, Stanford University, 382 Via Pueblo Mall, Stanford, CA 94305\\
$^{47}$Cerro Tololo Inter-American Observatory, Casilla 603, La Serena, Chile\\}

\appendix
\section{Study of mass-richness scatter for clusters found with the VT method}
\label{sec:vt_analysis}

The scatter of the mass-richness relation is a primary source of systematic uncertainties in cosmological measurements using galaxy clusters. To ensure that this and other astrophysical systematics are kept under control in future cosmology analyses, the DES collaboration has pursued the development of multiple cluster finding algorithms. Here we present initial results obtained using the analysis framework described throughout this paper on a DES-SVA1 cluster catalog created using the Voronoi Tessellation (VT) cluster finder \citep{soares11}. The VT method is fundamentally different from RM. Specifically, VT uses photometric redshifts to detect clusters in 2+1 dimensions, and is designed to produce a cluster catalog up to $z \sim 1$ and down to mass $M \sim 10^{13.5} \msun$ without any assumptions about the colours of galaxies in cluster environments. VT has been tested on DES simulations \citep{soares11} and on SDSS data. The mean mass-richness relation has been calibrated using a stacked weak lensing analysis of the SDSS VT clusters  \citep{2015arXiv150106893W}. In this appendix, we describe the first study of the scatter of the mass-richness relation using our analysis framework. This study demonstrates that VT, although not as mature a cluster finder as RM, is a promising algorithm. Our report provides an assessment of the current VT performance and helps to identify areas where improvements should be made. Some of these improvements can be applied to other cluster finding algorithms, too.

\subsection{VT method}
To detect clusters with the VT method, we build 2D tessellations  in each  photometric redshift shell and flag galaxies that lie in high-density cells as cluster members.  The density threshold is set in a non-parametric 
way from the 2-point correlation function of that given shell.  This takes advantage of the fact that the distribution of VT cell densities can be uniquely predicted for any given point process.  The 2-point function is a good description of the point process of the background galaxies on the sky.  Clusters cause a small deviation from the predicted distribution, and we take the point where that deviation is maximized as the threshold for detection.

\subsection{VT catalog for DES SVA1 data}
For the DES SVA1 data, the final catalog consists of 12948 clusters with richness $N_{vt} > 5$ and redshifts in the range $0.15 < z < 1$.  $N_{vt}$ is defined as the number of member galaxies.
The catalog covers the SPT-E and SPT-W regions of the SVA1 total footprint. We use DESDM data products as inputs, namely the \textit{Gold} galaxy catalog, plus photometric redshifts and mask information. We used a mask to apply magnitude cuts  $10 < $ mag\_auto\_i $ < 23.5$. The photometric redshift information 
was obtained using a neural network method \citep{sanchez14}.

We match the VT and SPT-SZ catalogs using the same method as described in Section 2.3. We sort the SPT-SZ cluster sample according to decreasing SPT observable $\xi$ and sort the VT catalogs according to decreasing richness. Then we associate the SPT cluster candidate with the richest  cluster candidate whose centre lies within 1.5 $R_{500}$ of the SPT-SZ centre. We finally remove the associated optical cluster from the list of possible counterparts when matching the remaining SPT-selected clusters. This procedure results in 42 VT clusters matched  to $\xi > 4.5$ SPT-SZ clusters.

\subsection{Results}

\begin{figure*}
\hbox to \hsize{
    \includegraphics[width=80mm]{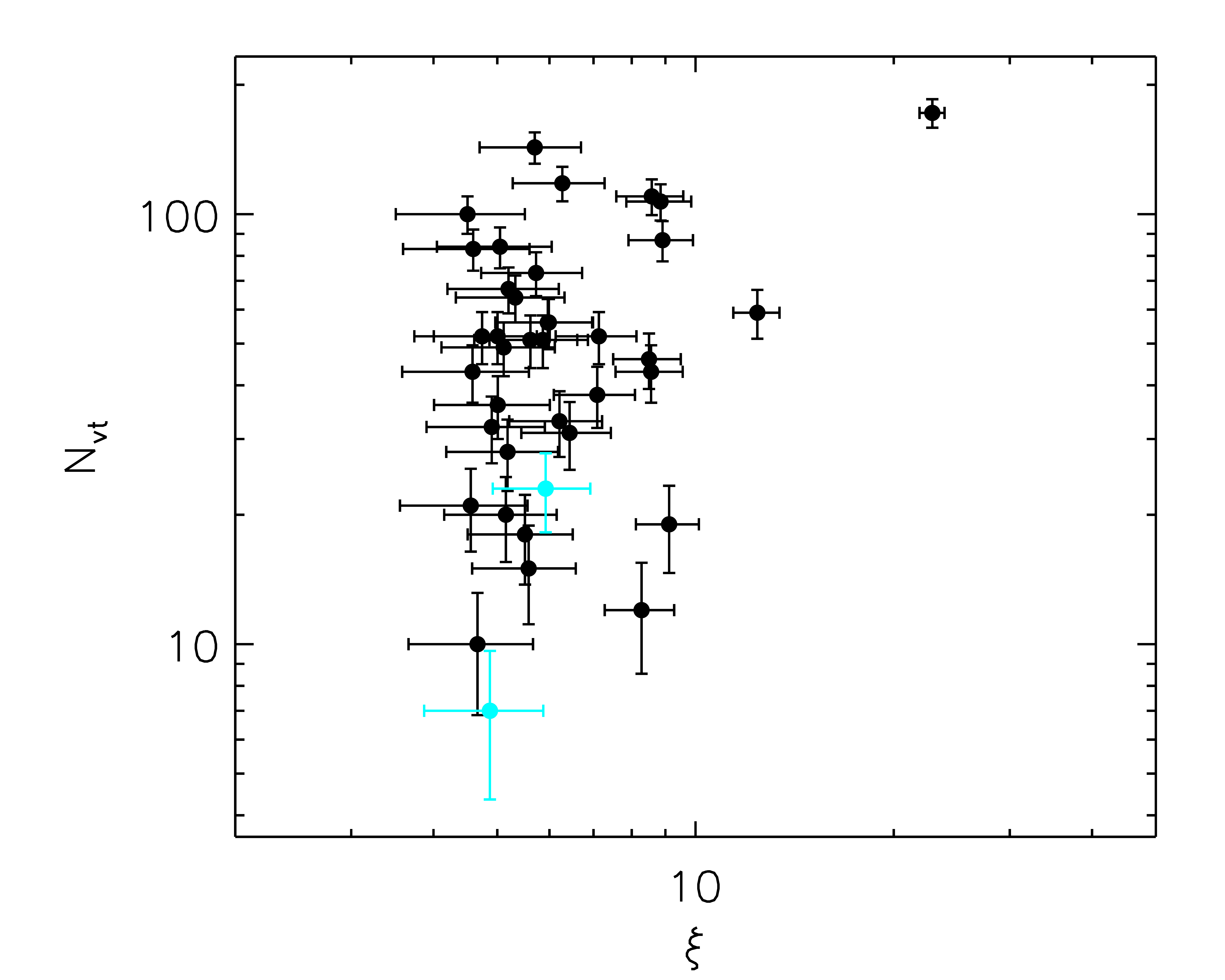}\hfil
    \includegraphics[width=80mm]{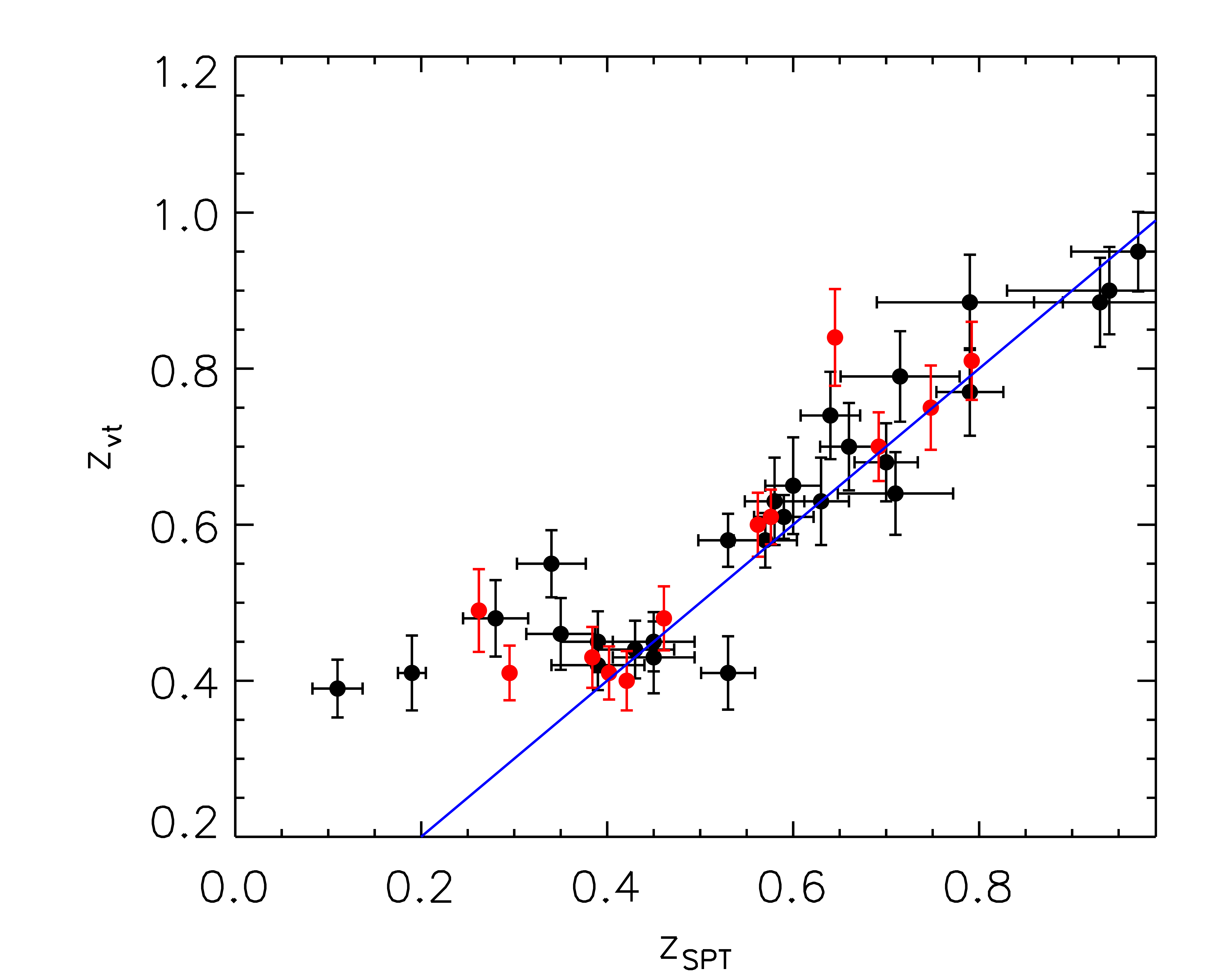}\hfil}
\caption{ \textit{Left panel:} Richness as a function of the SPT-SZ
  significance $\xi$ for the SPT-SZ+VT cluster sample. Clusters at $z<0.25$ are not not used for the richness-mass fit and are shown in cyan.  \textit{Right panel:} The estimated redshift for the VT sample as a function of SPT-SZ redshifts as presented in B15.  SPT-SZ candidates with spectroscopic redshift are shown in red.}
\label{vtplots}
\end{figure*}

In Figure~\ref{vtplots} we show the optical richness \nvt \, as a function of
$\xi$ (left panel) for the 42 matched clusters. The scatter in the
richness-$\xi$ plot indicates that the VT richness performs poorly as
a mass indicator. Improvements to the method are being developed based
on these findings. Specifically, future work will explore using the
galaxy magnitudes to calculate total stellar masses, which can then
be used as a mass proxy.  

Figure~\ref{vtplots} also shows the  VT estimated redshifts versus the redshifts determined in SPT follow-up observations (right panel). There is good agreement for $z>0.35$. Deviations at lower $z$ are found to arise from problems in the calculation of the 2-point function predictions at $z<0.35$, and a fix is underway.

We also obtained the best fit parameters, and the corresponding 68\% c.l.\ uncertainties,
for the richness-mass scaling relation described by Equations~\ref{eq:lambda_scaling} and \ref{eq:lambda_scatter}: 
\begin{eqnarray}
A_\textrm{VT}&=48.1^{+6.9}_{-6.3},\,      &B_\textrm{VT}=0.56^{+0.51}_{-0.25},\nonumber\\ 
C_\textrm{VT}&=-0.51^{+1.82}_{-1.48},\, &D_\textrm{VT}=0.64^{+0.29}_{-0.14}.
\label{eq:vtresults}
\end{eqnarray}
This result is consistent with the calibration for the mean relation obtained
with weak lensing \citep{2015arXiv150106893W}. The uncertainties, however, are larger than 
those obtained for RM. We expect that improvements 
to the richness estimator will result in better performance in future 
applications of the VT method. VT is not as mature a cluster finder
method as RM but the complementarity of the two techniques argues for further development of this alternative. 
This study allowed us to identify 
a key area for improvement and establishes a framework for future assessment
of the mass-richness scatter for VT clusters.

\end{document}